\documentclass[aps,prl,preprint,superscriptaddress,longbibliography]{revtex4-1}

\usepackage{graphicx}
\usepackage{color}
\usepackage{amsfonts}
\let\oldAA\AA
\renewcommand{\AA}{\text{\normalfont\oldAA}}

\usepackage{setspace}

\begin{document}
\title{Revealing flat bands and hybridization gaps in a twisted bilayer graphene device with microARPES} 
\author{Zhihao Jiang$^{\dagger}$}
\author{Kimberly Hsieh$^{\dagger}$}
\author{Alfred J. H. Jones}
\author{Paulina Majchrzak}
\author{Chakradhar Sahoo}
\affiliation{Department of Physics and Astronomy, Interdisciplinary Nanoscience Center, Aarhus University, 8000 Aarhus C, Denmark}
\author{Kenji~Watanabe}
\affiliation{Research Center for Electronic and Optical Materials, National Institute for Materials Science, 1-1 Namiki, Tsukuba 305-0044, Japan}
\author{Takashi~Taniguchi}
\affiliation{Research Center for Materials Nanoarchitectonics, National Institute for Materials Science,  1-1 Namiki, Tsukuba 305-0044, Japan}
\author{Jill A. Miwa}
\affiliation{Department of Physics and Astronomy, Interdisciplinary Nanoscience Center, Aarhus University, 8000 Aarhus C, Denmark}
\author{Yong P.  Chen}
\affiliation{Department of Physics and Astronomy, Interdisciplinary Nanoscience Center, Aarhus University, 8000 Aarhus C, Denmark}
\affiliation{Department of Physics and Astronomy and School of Electrical and Computer Engineering and Purdue Quantum Science and Engineering Institute, Purdue University, West Lafayette, IN 47907, USA \\$\dagger$ These authors contributed equally.  \\ Corresponding author email: ulstrup@phys.au.dk}
\author{S{\o}ren Ulstrup}
\email{ulstrup@phys.au.dk}
\affiliation{Department of Physics and Astronomy, Interdisciplinary Nanoscience Center, Aarhus University, 8000 Aarhus C, Denmark}

\begin{abstract}
Controlling the electronic structure of two-dimensional materials using the combination of twist angle and electrostatic doping is an effective means to induce emergent phenomena.  In bilayer graphene with an interlayer twist angle near the magic angle, the electronic dispersion is strongly modified by a manifold of hybridizing moir\'e Dirac cones leading to flat band segments with strong electronic correlations.  Numerous technical challenges arising from spatial inhomogeneity of interlayer interactions, twist angle and device functionality have so far limited momentum-resolved electronic structure measurements of these systems to static conditions.  Here,  we present a detailed characterization of the electronic structure exhibiting miniband dispersions for twisted bilayer graphene, near the magic angle, integrated in a functional device architecture using micro-focused angle-resolved photoemission spectroscopy.  The optimum conditions for visualizing the miniband dispersion are determined by exploiting the spatial resolution and photon energy tunability of the light source and applied to extract a hybridization gap size of $(0.14 \pm 0.03)$~eV and flat band segments extending across a moir\'e mini Brillouin zone.  \textit{In situ} electrostatic gating of the sample enables significant electron-doping,  causing the conduction band states to shift below the Fermi energy.  Our work emphasizes key challenges in probing the electronic structure of magic angle bilayer graphene devices and outlines conditions for exploring the doping-dependent evolution of the dispersion that underpins the ability to control many-body interactions in the material.\\

KEYWORDS: Moir\'e superlattice,  magic angle bilayer graphene, electronic structure, microARPES,  \textit{in situ} gating
\end{abstract}

\maketitle

\section{Introduction}

Heterostructures composed of two graphene layers stacked with a minute interlayer twist angle exhibit long-range moir\'e superlattices with strong hybridization effects between the linear Dirac bands of the constituent layers,  giving rise to van Hove singularities near the Fermi energy,  $E_F$ \cite{Lopes2007, Morell:2010,Guohong:2010,Bistritzer:2011,Santos:2012}.  Around the magic angle twist of 1.1$^{\circ}$,  a flat band develops as a result of hybridization between the Dirac bands at $E_F$,  leading to the emergence of several strongly correlated electronic phases \cite{Cao:2018b,Cao2018,Kerelsky:2019,Yonglong:2019,Jiang:2019,Choi:2019,Yankowitz:2019,Xiaobo:2019}.  The similarity of the associated temperature- and doping-dependent electronic phase diagram to that of the high-temperature cuprate superconductors has inspired the notion of applying twisted two-dimensional (2D) materials as solid state quantum simulators with the ability to tailor superlattices,  band structures and many-body interactions \cite{Andrei:2020,Kennes:2021}.  Angle-resolved photoemission spectroscopy (ARPES) has been an essential technique for uncovering the energy- and momentum-dependent electronic excitations that underpin the phase diagram of high-temperature superconductors \cite{Damascelli:2003,Sobota:2021}.  Employing this methodology to access the quasiparticle spectra of emerging correlated phases in twisted 2D materials is an important step to guide the design of interactions in superlattices.  Indeed,  the approach has led to the direct visualization of a flat band in twisted bilayer graphene (tBLG) near the magic angle \cite{Utama:2021,Lisi:2021,Yiwei:2022}.  However,  in order to gain access to energy- and momentum-dependent quasiparticle interactions that are relevant in transport experiments it is necessary to probe the electronic structure beyond the static conditions by applying voltages \textit{in situ} while performing the ARPES experiment.  

Such experiments on tBLG are extremely challenging for the following reasons: Firstly,  samples are typically produced using the dry-transfer method  \cite{Kyounghwan:2016,Kyounghwan:2017}, 
which is prone to hydrocarbon contamination from the fabrication process that ultimately lead to broadening of the ARPES spectra.  Hence,  annealing of the assembled stack at elevated temperatures is necessary to remove any adsorbates.  These processing steps naturally lead to inhomogeneities in twist angle,  strain and interlayer interactions that suppress the salient correlation effects \cite{Uri:2020,Lau:2022}.  It is therefore imperative to employ ARPES with nano- or micrometer spatial resolution (nanoARPES and microARPES).  Secondly,  \textit{in situ} control of charge carrier concentration is most elegantly achieved by integrating the twisted 2D material in a functional device architecture that permits electrostatic gating.  Such \textit{in operando} micro- and nanoARPES experiments have recently been demonstrated, but they remain highly challenging as the additional electrostatic field may impose detrimental energy- and momentum-broadening \cite{Joucken:2019,Nguyen:2019,muzzio2020,Jones2020,Hofmann:2021,Dale:2022,Jones:2022,Nunn:2023,nDale:2023}.  Moreover,  the induced photocurrent from the probing synchrotron light appears to strongly affect the device components, including the hexagonal boron nitride (hBN) dielectric that is customarily used,  causing gate leakage currents and degradation of device performance after prolonged exposure \cite{Velasco:2014}.  So far,  back gated ARPES measurements have been reported for single-layer graphene \cite{Nguyen:2019,muzzio2020,Dale:2022},  Bernal-stacked bilayer graphene \cite{Joucken:2019},  tBLG with a twist angle of 12.2$^{\circ}$ \cite{Jones2020},  tBLG with a twist angle around 3$^{\circ}$ \cite{nDale:2023} and twisted monolayer-on-bilayer graphene with a twist angle of 3.4$^{\circ}$ \cite{Nunn:2023},  but there are no such gated ARPES measurements reported for magic angle tBLG. 

Here,  we present a detailed microARPES study of a tBLG device with a twist angle around the magic angle performed at the SGM4 beamline of the ASTRID2 light source at Aarhus University,  Denmark \cite{Bianchi:2023,Volckaert:2023}.  We characterize the spatially-dependent electronic structure of the tBLG device and examine spectral intensity variations stemming from photon energy-dependent photoelectron interference effects.  The optimal combinations of sample spatial position and photon energy are identified and applied to extract hybridization gaps and determine the $(E,k)$-dependence of flat band segments in near-magic angle tBLG.  Finally, we present measurements during \textit{in situ} gating and discuss the impact of electrostatic doping on the electronic structure of near-magic angle tBLG.

\section{Results and discussion}

\begin{figure*}[t!] 
\begin{center}
\includegraphics[width=0.9\textwidth]{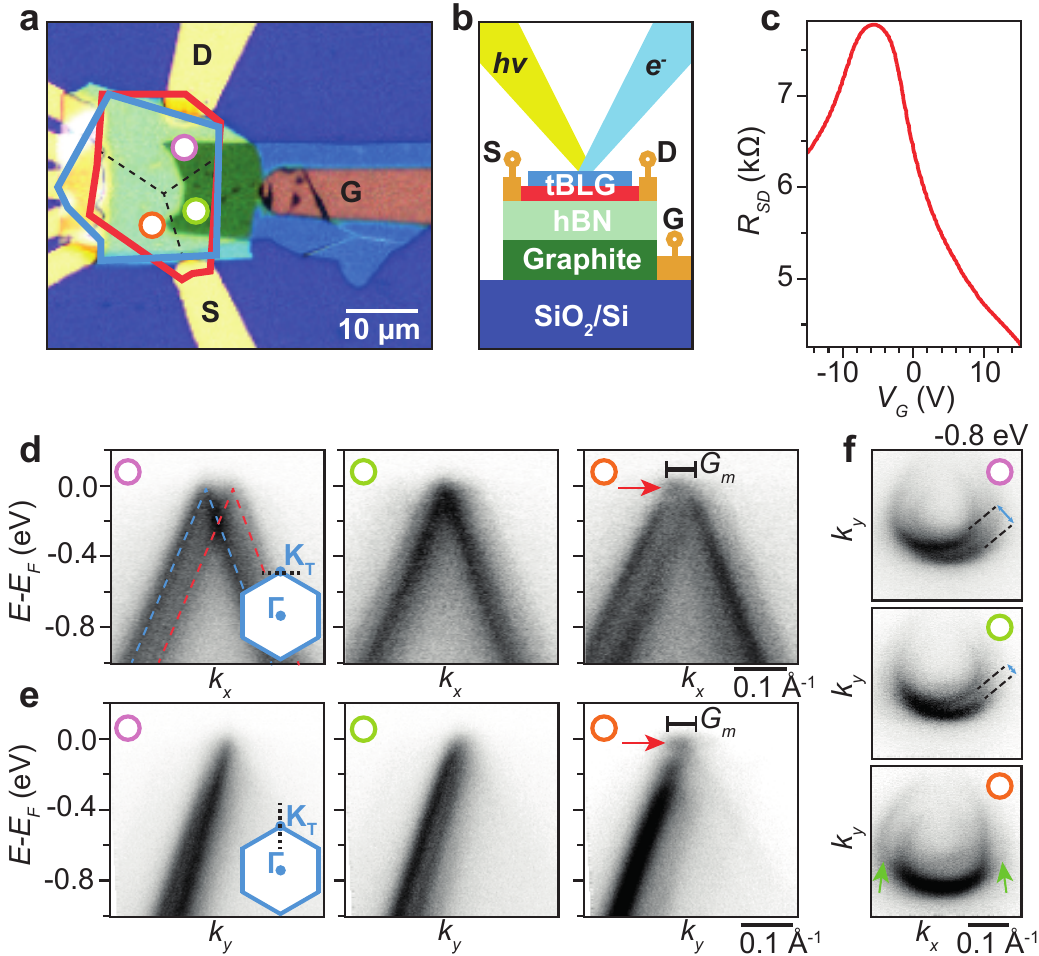}
\caption{Spatially-dependent electronic structure of tBLG device: (a) Optical micrograph of device with top (bottom) graphene layer demarcated by blue (red) outlines.  Dashed lines indicate regions of tBLG with different ARPES dispersions.  Source (S), drain (D) and gate (G) electrodes are indicated. The contacts visible in the left side of the image are disconnected.  (b) Schematic of micro-focused photoemission process and device architecture.  (c) Source-drain resistance as a function of gate voltage measured \text{in situ} after exposure to synchrotron light.  (d)-(e) ARPES spectra measured (d) perpendicular and (e) parallel to the $\mathrm{\Gamma}-\mathrm{K}_T$ direction, as indicated by the dashed lines in the blue BZ diagrams corresponding to the top graphene layer.  The spectra are obtained from the areas demarcated by correspondingly colored circles in (a).  The blue and red dashed lines are outlines of top and bottom graphene Dirac cones,  respectively. The red arrows indicate the flat band expected around the magic angle.  The moir\'e lattice vector length,  $G_m$,  is demarcated by a bracket for a twist angle of 1.2$^{\circ}$.  (f) Constant energy cuts extracted at -0.8 eV from the same areas.  The dashed lines separated by double-headed blue arrows indicate rigid $k$-shifts of the energy contours.  The green arrows demarcate dim lobes from interacting graphene layers. The spectra were acquired at a photon energy $h\nu$ of 47 eV.}
\label{fig:1}
\end{center}
\end{figure*}

Our device is composed of tBLG with a nominal twist angle of 1.2$^{\circ}$ targeted during the fabrication.  The stack is supported on hBN, with a thickness of 30 nm, that makes a twist angle of $(2.2 \pm 0.4)^{\circ}$ with respect to the top graphene layer of the tBLG,  as determined from the orientation of the Brillouin zone (BZ) in the ARPES measurements \cite{Koch:2018ab}.  A graphite back gate below the hBN partially overlaps the area of the tBLG.  This configuration is chosen to  allow for electrostatic gating of the tBLG. The device is placed on 285 nm SiO$_2$/$p$-doped Si.  A three-terminal configuration is used with Au/Cr source and drain contacts to the bottom layer of the tBLG stack and a contact to the graphite flake.  An optical micrograph of the device is shown in Fig.  \ref{fig:1}(a),  and a schematic of the architecture and photoemission process is presented in Fig.  \ref{fig:1}(b).  Photons with tunable energy $h\nu$ are focused to a spot with minimum lateral diameter of $4.4$ $\mu$m in the present experiment by using an achromatic capillary mirror \cite{Koch2018,Ulstrup2020,Volckaert:2023}.  The kinetic energy and angular distributions of photoemitted electrons are measured using a SPECS Phoibos 150 SAL hemispherical analyzer.  In order to verify that the electrodes are functional,  a sweep of the source-drain resistance, $R_{SD}$, versus gate voltage, $V_G$, is performed,  as shown in Fig.  \ref{fig:1}(c).  The resistance reaches a maximum of 7.8 k$\Omega$ at the gate voltage where the carrier density has a minimum.  The shape of the $R_{SD}(V_G)$-curve is consistent with earlier transport measurements of single-, bi- and multi-layer graphene devices \cite{Morozov:2008,Yang:2011,Maassen:2011},  indicating that our device is operational.  Note that the resistance sweep was done after exposure to the synchrotron beam, which may influence the resistance peak position and width \cite{muzzio2020}.

Spatially-resolved measurements of the $(E,k)$-dependent photoemission intensity over the area shown in the optical micrograph reveal that the tBLG part exhibits three distinct types of dispersions that are found in the regions delineated by dashed lines in Fig.  \ref{fig:1}(a).  Example ARPES spectra from these three regions are presented in Figs.  \ref{fig:1}(d)-(e) and color-coded via circles that refer to the spots marked in the optical image. The spectra represent cuts perpendicular and parallel to the $\mathrm{\Gamma}-\mathrm{K}_T$ direction, where $\mathrm{K}_T$ is the Dirac point of the top graphene layer.  Corresponding $(k_x,k_y)$-dependent constant energy cuts at -0.8 eV are shown in  Fig.  \ref{fig:1}(f).  This energy is chosen as it is simpler to distinguish the dispersion emerging from two cones than closer to $E_F$. The intensity variation in these contours mainly derives from sublattice interference effects between photoelectrons emitted from the graphene layers.  This causes a $k$-dependent modulation of the photoemission matrix elements known as the dark corridor in graphene \cite{Gierz2011.PhysRevB.83.121408},  which is characterized by a high intensity towards the center of the first BZ and a suppression in the direction towards the neighbouring BZs.  Notably,  this causes a strong suppression of intensity from half of the dispersion in the cuts along $\mathrm{\Gamma}-\mathrm{K}_T$ as seen in Fig.  \ref{fig:1}(e).  

A more detailed inspection of the spectra from the three regions reveals several key differences between them.  The region marked by a purple circle displays two sets of Dirac cones that cross without any sign of interaction between their bands.  The cone outlined by blue dashed lines is approximately twice as intense as the cone outlined by red dashed lines,  which enables us to assign the cone outlined by blue dashed lines to the top graphene layer \cite{Jones2020}.  Analysis of the constant energy contour reveals a momentum shift $\Delta k$ of $(0.06~\pm~0.01)$~$\AA^{-1}$ between the two cones,  which is demarcated by a blue double-headed arrow in Fig.  \ref{fig:1}(f).  The twist angle $\theta$ is then calculated to be $(2.0 \pm 0.2)^{\circ}$ using $\sin(\theta / 2) = \Delta k/2 |\mathrm{K} |$,  where $|\mathrm{K} | = 1.7$~$\AA^{-1}$ is the distance from $\mathrm{\Gamma}$ to $\mathrm{K}$ of graphene.  The region marked by a green circle also displays two non-interacting Dirac cones that are shifted even less in momentum with respect to each other. This is most clearly seen via the two closely spaced linear branches in Fig.  \ref{fig:1}(e) and slightly shifted contours in Fig. \ref{fig:1}(f).  The shift is determined to be $\Delta k = (0.04 \pm 0.01)$ $\AA^{-1}$, which corresponds to a twist angle of $(1.4 \pm 0.2)^{\circ}$.  In the region marked by an orange circle an immediately striking feature resembling a flat band is noticeable, as seen via the red arrows in Figs. \ref{fig:1}(d)-(e).  The constant energy contour exhibits additional segments of intensity, demarcated by green arrows,  that cannot be described as just two shifted single-layer graphene contours, as in the other regions. These spatially-resolved ARPES spectra establish the region of the device where the two graphene layers interact.  The extent of the flat band segment is comparable to the moir\'e lattice vector around the magic angle twist, which is given by $G_m = 0.06$ $\AA^{-1}$,  as indicated by the brackets in Figs.  \ref{fig:1}(d)-(e).  The features are thus consistent with those expected from tBLG near the magic angle \cite{Utama:2021,Lisi:2021,Yiwei:2022}.  The non-interacting parts likely contain graphene layers that are not sufficiently mechanically pressed together during stacking. It is also possible that residual strain,  impurities between layers and pits in the underlying hBN spoil the contact between the flakes.  In the following, we focus solely on the dispersion from the interacting region of the device.

\begin{figure*}[t!] 
\begin{center}
\includegraphics[width=0.9\textwidth]{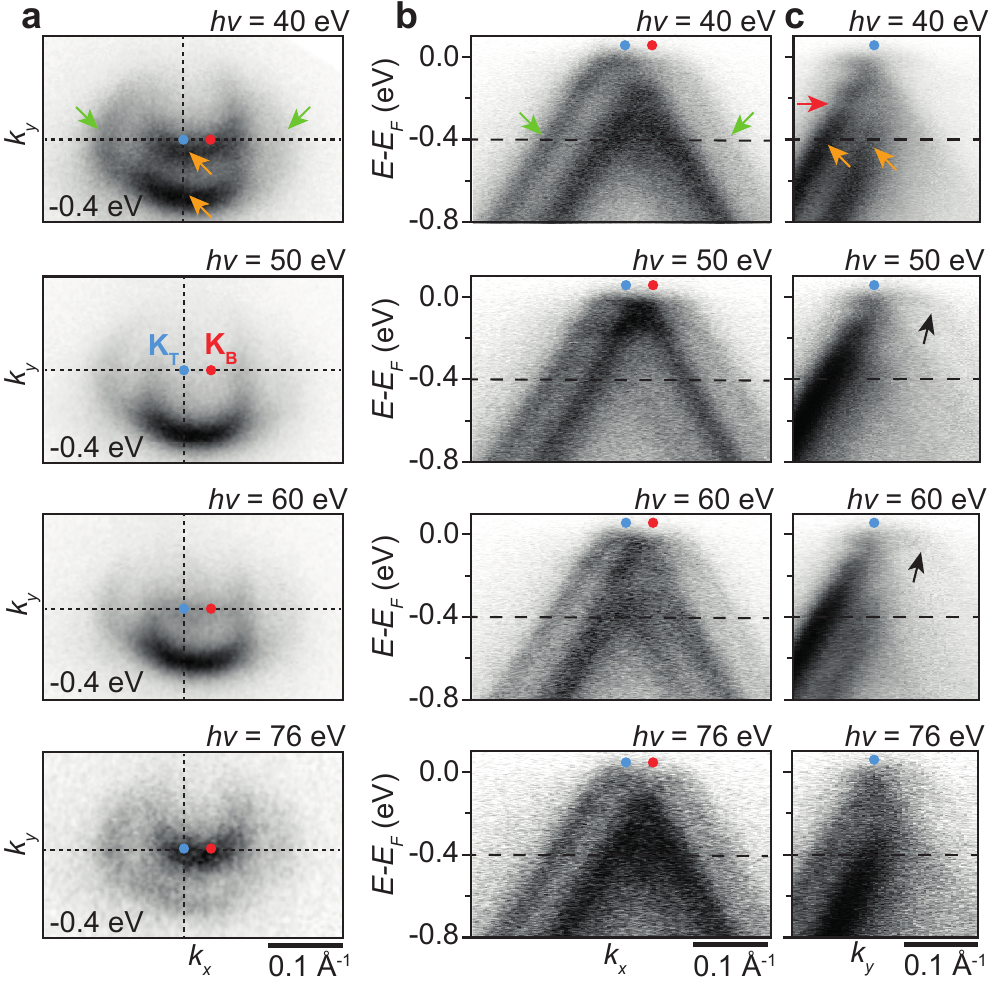}
\caption{Photon energy dependence of tBLG ARPES intensity around the magic angle: (a) Constant energy cuts at -0.4 eV at the given values of $h\nu$.  (b)-(c) ARPES spectra extracted  along  (b) horizontal and (c) vertical dotted lines  in (a).  The dashed lines in (b)-(c) indicate the energy where the cuts in (a) are obtained.  The photon energy is the same as stated in (a) along the panel rows. The location of $\mathrm{K}_B$- and $\mathrm{K}_T$-points are marked by red and blue dots, respectively.  Green and orange arrows link side lobes and arcs in the constant energy cuts to dispersing branches in the $E(k)$-cuts.  Black arrows indicate minibands along the dark corridor.  The red arrow demarcates a hybridization gap around -0.2 eV. }
\label{fig:2}
\end{center}
\end{figure*}

A moir\'e superlattice arises from the twisting of the graphene layers, which appears as minibands in the electronic structure measurements. The interactions between the minibands give rise to hybridization gaps that are appreciable in the photoemission intensity. To further substantiate the previous analysis, we proceed to search for the optimum photon energy where the hybridization effects are most visible.  Such a dependence on photon energy is expected due to interference effects involving photoelectrons from the moir\'e sites,  in addition to the two sublattice sites in each graphene layer \cite{Lisi:2021,Nunn:2023}.  By performing a scan of $h\nu$ from 40 to 76 eV we are able to determine substantial redistributions of intensity between the manifold of bands in tBLG near the magic angle, as shown in Fig. \ref{fig:2}.  Representative spectra have been selected at photon energies of 40,  50,  60 and 76 eV, as these highlight the most dramatic intensity changes within the probed range.  

Figure \ref{fig:2}(a) presents constant energy contours at -0.4 eV around the $\mathrm{K}_B$- and $\mathrm{K}_T$-points, corresponding to the main BZ $\mathrm{K}$-points of bottom and top graphene layers,  respectively.  In this energy cut, the features resulting from the superlattice are clearly distinguishable,  whereas at lower energies the hybridization effects and flat band lead to a larger degree of complexity.  At a photon energy of 40 eV,  two arcs of nearly equal intensity occur (see orange arrows) along the vertical dotted line through $\mathrm{K}_T$,  in addition to a set of dim lobes (see green arrows) on the sides of $\mathrm{K}_B$ and $\mathrm{K}_T$ along the horizontal dotted line.  From 50 to 76 eV, the intensity switches from the outer arc to the inner arc centred on the $\mathrm{K}_B$-$\mathrm{K}_T$ line,  while the side lobes remain relatively dim.  $E(k)$-spectra resulting from horizontal and vertical cuts along the dotted lines are displayed in Figs.  \ref{fig:2}(b)-(c).  The two side lobes are formed by two linearly dispersing bands (see green arrows in  Fig.  \ref{fig:2}(b)) that disperse from lower energy and hybridize to form a flat band around $E_F$.  Cuts along the vertical direction similarly exhibit two linearly dispersing bands (see orange arrows Fig.  \ref{fig:2}(c)) that form the arcs seen in the constant energy contours.  The linear parts are interrupted by the formation of hybridization gaps around -0.2 eV (see red arrow in Fig.  \ref{fig:2}(c)). The relative intensity levels of these branches vary significantly.  At 50 eV, the intensity of the outer branch is high while it is almost completely suppressed for the inner branch.  At 76 eV, this behavior is reversed.  The important consequence of these intensity variations is that faint minibands that form an extended flat band at $E_F$ are resolvable in the dark corridor side of the vertical cut at 50 and 60 eV (see black arrows in Fig.  \ref{fig:2}(c)).  Additionally,  hybridization effects are most pronounced at 40 eV where the relative intensity levels are more similar between the branches,  as exemplified by the gap highlighted by a red arrow in Fig.  \ref{fig:2}(c).  We therefore choose a photon energy of 40 eV for more in-depth analysis of the hybridization effects and the extent of the flat band dispersion, as discussed in the following.

\begin{figure*}[t!] 
\begin{center}
\includegraphics[width=1.0\textwidth]{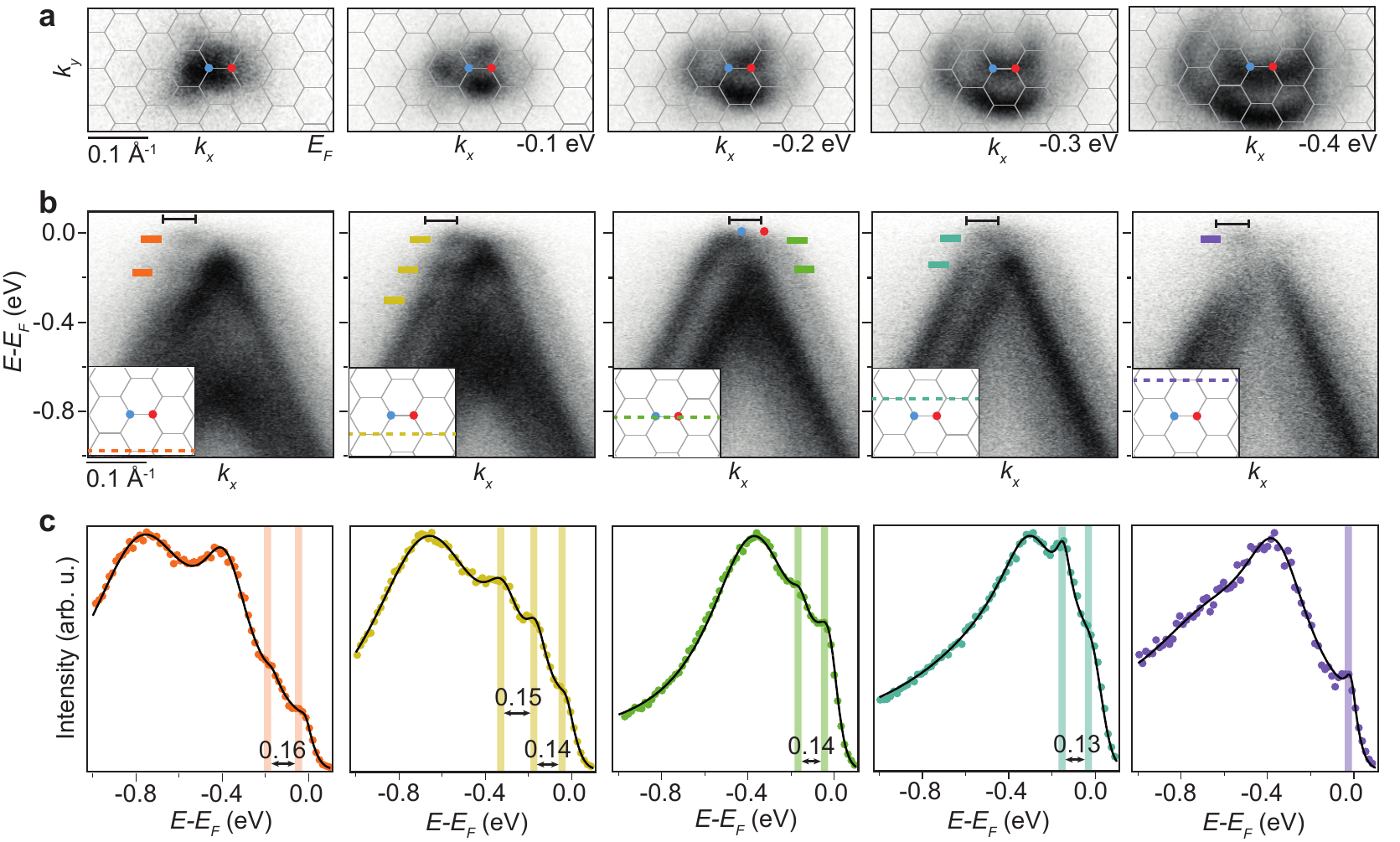}
\caption{Tracking hybridization effects across mini Brillouin zones: (a) Constant energy cuts extracted at the given energies.  The mBZs corresponding to a moir\'e lattice vector of 0.06 $\AA^{-1}$ have been overlaid.  Red and blue dots indicate $\mathrm{K}_B$ and $\mathrm{K}_T$, respectively.  (b) ARPES spectra extracted along mBZ high symmetry directions as indicated by dashed lines in the corresponding mBZ diagrams.  (c) Energy distribution curves (markers) integrated over a $k$-range of 0.05 \AA$^{-1}$. The integration region is indicated by a bracket in the corresponding ARPES spectrum in the same column in (b).  Peak-to-peak energy separations extracted from fits (black curves) are indicated by vertical bars separated by double-headed arrows and values are stated in units of eV.  The error bar on the extracted energies is $\pm 0.03$ eV. The fitted peak positions are shown via ticks in (b).  The ARPES spectra were obtained at $h\nu = 40$ eV.}
\label{fig:3}
\end{center}
\end{figure*}

The evolution of constant energy contours from $E_F$ to -0.4 eV with overlaid mini Brillouin zones (mBZs) corresponding to a near-magic angle moir\'e lattice vector $G_m$ of 0.06 $\AA^{-1}$ is presented in Fig.  \ref{fig:3}(a).  Around $E_F$,  the intensity is primarily concentrated in the four mBZs surrounding $\mathrm{K}_B$ and $\mathrm{K}_T$, consistent with previous reports \cite{Lisi:2021,Utama:2021}.  Horizontal high symmetry $E(k)$-cuts along these mBZs are shown in Fig.  \ref{fig:3}(b).  We observe flat bands that are separated by gaps at the energies noted by tick marks.  The features are analyzed in Fig.  \ref{fig:3}(c) using energy distribution curves (EDCs) integrated over a $k$-range of 0.05 $\AA^{-1}$ centred on the flat band segments,  as shown via brackets in Fig.  \ref{fig:3}(b).  The dispersive bands give rise to broad peaks in the EDCs from -0.4 eV and towards lower energies.  Step-like features marked by vertical bars between $E_F$ and -0.4 eV correspond to flat band segments separated by hybridization gaps.  The EDCs are modelled by Lorentzian peaks multiplied by a Fermi-Dirac cut-off.  Fitting to this model reveals a peak pinned at $(-0.04 \pm 0.03)$ eV in all of the mBZ cuts,  consistent with the flat band expected for tBLG near the magic angle \cite{Lisi:2021,Utama:2021}.  The average linewidth of the flat band determined from the full width at half maximum (FWHM) of the Lorentzian fits is $(0.07 \pm 0.02)$ eV.  This energy broadening is likely attributed to sub-micrometer inhomogeneities of the twist angle arising from strain fluctuations and stacking faults \cite{Uri:2020,Lau:2022}. The average value of the associated hybridization gaps is determined to be $(0.14 \pm 0.03)$ eV,  which reflects the interlayer interaction strength and moir\'e potential established between the twisted graphene layers.  The observed separation of the flat state from dispersive features by a hybridization gap is consistent with the theoretical results corroborated by experimental data in Refs.  \cite{Utama:2021,Lisi:2021}.  The  previously calculated values of the gap around 40 meV \cite{Utama:2021} and 110 meV \cite{Lisi:2021} appear to underestimate the interaction strength we observe.  Indeed,  the magnitude of the gap depends sensitively on lattice relaxation effects and is therefore difficult to model \cite{Utama:2021}.

\begin{figure*}[t!] 
\begin{center}
\includegraphics[width=0.5\textwidth]{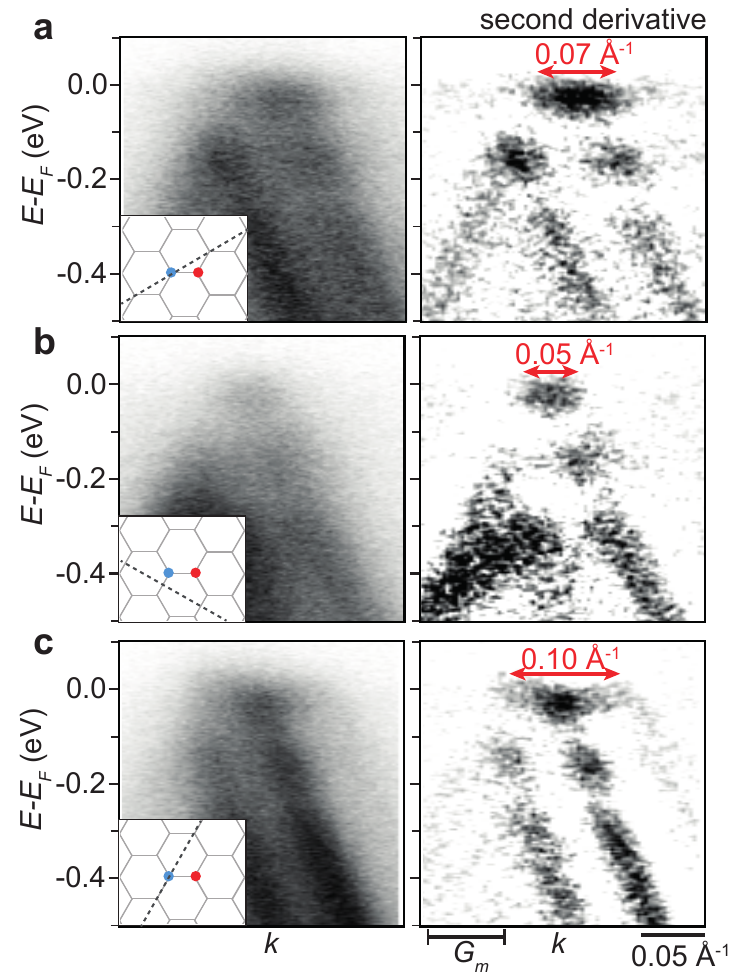}
\caption{Characterization of flat band dispersion: (a) ARPES intensity (left panel) and corresponding second derivative map of the intensity (right panel) measured along the cut marked by a dashed line in the mBZ diagram.  (b)-(c) Similar plots for spectra obtained along the directions indicated by dashed lines in the mBZ diagrams.  The estimated extent of the flat band in $k$ along the extracted cuts is indicated by red double-headed arrows on the second derivative maps.  The size of the moir\'e lattice vector $G_m$ is provided via a bracket for a twist angle of 1.2$^{\circ}$.  The measurements were performed at $h\nu = 40$ eV.}
\label{fig:4}
\end{center}
\end{figure*}

The intensity of minibands rapidly drops towards higher order mBZs due to the incommensurability of a general twist angle.  In other words,  the band structure of the superlattice is not a well-defined quantity such that the flat band dispersion acquires a finite momentum range in the measured ARPES spectral weight.  This range is visualized and extracted using the analysis presented in Fig.  \ref{fig:4}.  Three cuts are extracted along directions marked in the corresponding mBZ diagrams.  The second derivative of the ARPES intensity corresponding to these cuts is then determined \cite{CurvM:2011}.  The resulting second derivative images seen in Fig.  \ref{fig:4} emphasize flat band segments that are separated from the dispersive features by hybridization gaps.  The average extent of the flat band in momentum obtained from these cuts is $(0.07 \pm 0.03)$ $\AA^{-1}$,  which is in line with the moir\'e lattice vector of $G_m = 0.06$ $\AA^{-1}$, indicating coherence across roughly one moir\'e unit cell.

\begin{figure*}[t!] 
\begin{center}
\includegraphics[width=0.5\textwidth]{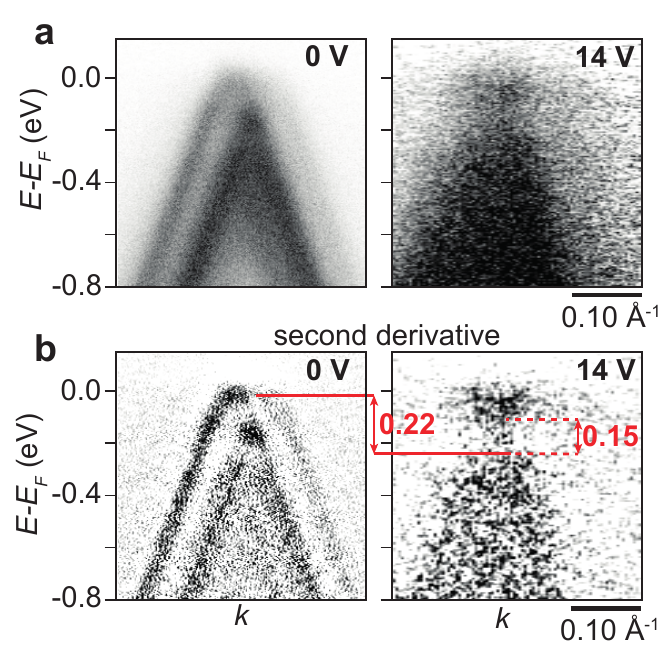}
\caption{Effect of \textit{in situ} electrostatic gating: (a) ARPES spectra extracted along the $\mathrm{K}_T-\mathrm{K}_B$ direction at 0 V (left panel) and 14 V (right panel).  (b) Corresponding plots of the second derivative of the intensity.  Red lines indicate a shift of the valence band manifold.  Dashed red lines demarcate an apparent gap in the dispersion measured at 14 V.  The corresponding energies are marked by double-headed red arrows and stated in units of eV.  The error bars are $\pm 0.05$ eV.  The spectra were measured at $h\nu = 40$ eV.}
\label{fig:5}
\end{center}
\end{figure*}

Finally,  we utilize the above analysis in order to investigate the impact on the ARPES dispersion of applying an electrostatic gate voltage of 14 V while keeping the tBLG grounded.  This is the maximum voltage we could achieve while avoiding a large gate leakage current.  ARPES spectra obtained along $\mathrm{K}_T-\mathrm{K}_B$ are compared at 0 and 14 V in Fig.   \ref{fig:5}(a).  The main effect of the applied electric field is a substantial momentum broadening,  which precluded more detailed measurements at intermediate gate voltages as the gradual changes in the spectra were not possible to discern.  Consequently, we are limited to estimating  the doping-induced energy shift of the tBLG bands,  which is done via the second derivative intensity in Fig.   \ref{fig:5}(b).  The broadened manifold of valence bands shifts down by $(0.22~\pm~0.05)$~eV,  revealing an apparant gap of $(0.15~\pm~0.05)$~eV and a distribution of states around $E_F$ that likely corresponds to the conduction band of tBLG.  Such a gap may develop due to the Coulomb potential energy difference between the two graphene layers induced by the perpendicular electric field,  as seen in potassium-doped Bernal-stacked bilayer graphene \cite{Ohta:2006}.  Note that our data does not enable us to preclude the presence of faint in-gap states. Furthermore,  the electric field leads to different charge carrier concentrations in the top and bottom graphene layers (see Refs.  \cite{Jones2020,Nunn:2023,nDale:2023} and Methods Section),  which results in an increasingly asymmetric overlap of the top and bottom Dirac cones with voltage and thereby substantially affect the hybridization energies and possible flat band dispersion \cite{Lopes2007}.  These details are difficult to resolve due to the broadening of our spectra.  This is caused by the imperfect overlap of graphite back gate and tBLG, which is evident from the optical micrograph in Fig.  \ref{fig:1}(a). The resulting inhomogeneous electric field between back gate and the tBLG region marked by an orange circle in Fig.  \ref{fig:1}(a) causes broadening of the photoelectron momentum distributions, spoiling the linewidths of the acquired spectra.  Nevertheless,  our data implies a significant modification of the dispersion over the achievable range of gating.  Detailed $(E,k)$-dependent measurements and analysis of such effects as a function of gate voltage are critical to access and tailor the electronic structure of tBLG at small twist angles,  requiring further optimization of device designs for micro- and nanoARPES experiments.

\section{Conclusions}

We have applied microARPES to map the energy- and momentum-resolved dispersion of a twisted bilayer graphene device, allowing  us to spatially disentangle regions of the device where the graphene layers are interacting.  Variations in photoemission intensity are observed between minibands, which arise from photoelectron interference effects,  enabling us to select an optimum photon energy to visualize the dispersion and analyse hybridization effects. The dispersion in the interacting regions of the device is thereby found to be consistent with near-magic angle twisted bilayer graphene, exhibiting hybridization gaps of $(0.14~\pm~0.03)$~eV and flat band segments extending across a moir\'e mini Brillouin zone.  Electrostatic doping via an applied gate voltage causes a downwards shift of the manifold of interacting minibands,  while a narrow distribution of states appears around $E_F$ which is separated from the manifold by $(0.15~\pm~0.05)$~eV.  Our detailed microARPES analysis of a twisted bilayer graphene device highlights key challenges that are important to address in order to access energy- and momentum-resolved quasiparticle dynamics in the salient correlated phases of twisted 2D materials.

\section{Methods}

\textit{Device fabrication} \\
High-quality flakes of single-layer graphene,  hBN and graphite were exfoliated onto SiO$_{2}$/Si substrates and identified using an optical microscope.  The flakes were assembled into a heterostructure using the dry-transfer technique  \cite{Kyounghwan:2016,Kyounghwan:2017}.  We used a polycarbonate (PC)/polydimethylsiloxane (PDMS) stamp on a glass slide to pick up all flakes. Single-layer graphene was cut in two pieces using a tungsten micro-needle tip with a diameter of 0.001 mm.  One of the pieces was picked up at 130 $^{\circ}$C.  The other piece was rotated by $1.2^{\circ}$ and then picked up using the other piece.  The hBN flake was subsequently picked up at 120 $^{\circ}$C followed by the graphite flake at 130 $^{\circ}$C. The entire stack was then released onto a SiO$_{2}$/Si substrate with a pre-defined Au/Cr pattern with a thickness of $50/5$ nm at 180 $^{\circ}$C.  The device was finally annealed in a H$_{2}$/Ar environment at 250 $^{\circ}$C for $4$ hours and then wire-bonded to a chip carrier.\\

\textit{Photoemission measurements} \\
The microARPES measurements were performed at the SGM4 beamline of the ASTRID2 synchroton at Aarhus University, Denmark. The chip carrier with the device was annealed in the ultrahigh vacuum system at a temperature of 200 $^{\circ}$C for several hours before exposure to the beam. The device was kept at room temperature at a base pressure better than $3\cdot10^{-10}$~mbar during data acquisition.

The synchrotron beam was focused to a spot size of 4.4~$\mu$m using an elliptical capillary mirror [Sigray Inc.].  The focus and position of the spot was stable over the probed photon energy range from 40 to 76 eV. The ARPES spectra were obtained using a SPECS Phoibos 150 SAL analyser. The energy- and angular resolution were better than 20~meV and 0.1$^{\circ}$, respectively. The scans of $(E, k_x, k_y)$-dependent photoemission intensity were acquired with the scanning angle lens mode of the analyser while keeping the sample position fixed.  Photon energy dependent scans were done by collecting $(E, k_x, k_y)$-spectra at each photon energy in order to track the positions of the Dirac points. The sample was aligned such that the analyzer cut was along the direction of momentum space that is perpendicular to the $\mathrm{\Gamma}$-$\mathrm{K}_T$ high-symmetry direction of the top graphene layer.  

Electrostatic gating and source-drain resistance measurements were achieved using two Keithley 2450 sourcemeters.  It is not possible to directly determine the induced carrier concentration $n$ in the two layers from the data,  however,  one may obtain a qualitative estimate using the $n(V_G)$ dependence measured for the tBLG device with a twist angle of 12.2$^{\circ}$ investigated in Ref.  \citenum{Jones2020}.  The larger twist angle prevents any interaction between the linearly dispersing Dirac bands close to $E_F$,  enabling a direct extraction of $n = k_F^2/\pi$, where $k_F$ is the Fermi wavevector obtained from the ARPES spectra.  For tBLG near the magic angle this method does not work, as the dispersion is strongly modified via hybridization and flat band formation.  Based on the 12.2$^{\circ}$ tBLG device,  the doping in the bottom layer at 14 V would amount to $7 \cdot 10^{12}$ cm$^{-2}$ and in the top layer it would be $3 \cdot 10^{12}$  cm$^{-2}$ \cite{Jones2020}.

\section{acknowledgement}
The authors acknowledge funding from the Danish Council for Independent Research, Natural Sciences under the Sapere Aude program (Grant Nos.  DFF-9064-00057B and DFF-6108-00409),  the Aarhus University Research Foundation, the Novo Nordisk Foundation (Project Grant NNF22OC0079960) and from VILLUM FONDEN under the Villum Investigator Program (Grant. No. 25931).  C.S.  acknowledges Marie Sklodowska-Curie Postdoctoral Fellowship (proposal number 101059528).  Growth of hexagonal boron nitride was supported by the JSPS KAKENHI (Grant Numbers 20H00354, 21H05233 and 23H02052) and World Premier International Research Center Initiative (WPI), MEXT, Japan.


\begin{thebibliography}{45}%
\makeatletter
\providecommand \@ifxundefined [1]{%
 \@ifx{#1\undefined}
}%
\providecommand \@ifnum [1]{%
 \ifnum #1\expandafter \@firstoftwo
 \else \expandafter \@secondoftwo
 \fi
}%
\providecommand \@ifx [1]{%
 \ifx #1\expandafter \@firstoftwo
 \else \expandafter \@secondoftwo
 \fi
}%
\providecommand \natexlab [1]{#1}%
\providecommand \enquote  [1]{``#1''}%
\providecommand \bibnamefont  [1]{#1}%
\providecommand \bibfnamefont [1]{#1}%
\providecommand \citenamefont [1]{#1}%
\providecommand \href@noop [0]{\@secondoftwo}%
\providecommand \href [0]{\begingroup \@sanitize@url \@href}%
\providecommand \@href[1]{\@@startlink{#1}\@@href}%
\providecommand \@@href[1]{\endgroup#1\@@endlink}%
\providecommand \@sanitize@url [0]{\catcode `\\12\catcode `\$12\catcode
  `\&12\catcode `\#12\catcode `\^12\catcode `\_12\catcode `\%12\relax}%
\providecommand \@@startlink[1]{}%
\providecommand \@@endlink[0]{}%
\providecommand \url  [0]{\begingroup\@sanitize@url \@url }%
\providecommand \@url [1]{\endgroup\@href {#1}{\urlprefix }}%
\providecommand \urlprefix  [0]{URL }%
\providecommand \Eprint [0]{\href }%
\providecommand \doibase [0]{http://dx.doi.org/}%
\providecommand \selectlanguage [0]{\@gobble}%
\providecommand \bibinfo  [0]{\@secondoftwo}%
\providecommand \bibfield  [0]{\@secondoftwo}%
\providecommand \translation [1]{[#1]}%
\providecommand \BibitemOpen [0]{}%
\providecommand \bibitemStop [0]{}%
\providecommand \bibitemNoStop [0]{.\EOS\space}%
\providecommand \EOS [0]{\spacefactor3000\relax}%
\providecommand \BibitemShut  [1]{\csname bibitem#1\endcsname}%
\let\auto@bib@innerbib\@empty
\bibitem [{\citenamefont {Lopes~dos Santos}\ \emph {et~al.}(2007)\citenamefont
  {Lopes~dos Santos}, \citenamefont {Peres},\ and\ \citenamefont
  {Castro~Neto}}]{Lopes2007}%
  \BibitemOpen
  \bibfield  {author} {\bibinfo {author} {\bibfnamefont {J.~M.~B.}\
  \bibnamefont {Lopes~dos Santos}}, \bibinfo {author} {\bibfnamefont
  {N.~M.~R.}\ \bibnamefont {Peres}}, \ and\ \bibinfo {author} {\bibfnamefont
  {A.~H.}\ \bibnamefont {Castro~Neto}},\ }\bibfield  {title} {\enquote
  {\bibinfo {title} {Graphene bilayer with a twist: Electronic structure},}\
  }\href {\doibase 10.1103/PhysRevLett.99.256802} {\bibfield  {journal}
  {\bibinfo  {journal} {Phys. Rev. Lett.}\ }\textbf {\bibinfo {volume} {99}},\
  \bibinfo {pages} {256802} (\bibinfo {year} {2007})}\BibitemShut {NoStop}%
\bibitem [{\citenamefont {Su\'arez~Morell}\ \emph {et~al.}(2010)\citenamefont
  {Su\'arez~Morell}, \citenamefont {Correa}, \citenamefont {Vargas},
  \citenamefont {Pacheco},\ and\ \citenamefont {Barticevic}}]{Morell:2010}%
  \BibitemOpen
  \bibfield  {author} {\bibinfo {author} {\bibfnamefont {E.}~\bibnamefont
  {Su\'arez~Morell}}, \bibinfo {author} {\bibfnamefont {J.~D.}\ \bibnamefont
  {Correa}}, \bibinfo {author} {\bibfnamefont {P.}~\bibnamefont {Vargas}},
  \bibinfo {author} {\bibfnamefont {M.}~\bibnamefont {Pacheco}}, \ and\
  \bibinfo {author} {\bibfnamefont {Z.}~\bibnamefont {Barticevic}},\ }\bibfield
   {title} {\enquote {\bibinfo {title} {Flat bands in slightly twisted bilayer
  graphene: Tight-binding calculations},}\ }\href {\doibase
  10.1103/PhysRevB.82.121407} {\bibfield  {journal} {\bibinfo  {journal} {Phys.
  Rev. B}\ }\textbf {\bibinfo {volume} {82}},\ \bibinfo {pages} {121407}
  (\bibinfo {year} {2010})}\BibitemShut {NoStop}%
\bibitem [{\citenamefont {Li}\ \emph {et~al.}(2010)\citenamefont {Li},
  \citenamefont {Luican}, \citenamefont {Lopes~dos Santos}, \citenamefont
  {Castro~Neto}, \citenamefont {Reina}, \citenamefont {Kong},\ and\
  \citenamefont {Andrei}}]{Guohong:2010}%
  \BibitemOpen
  \bibfield  {author} {\bibinfo {author} {\bibfnamefont {Guohong}\ \bibnamefont
  {Li}}, \bibinfo {author} {\bibfnamefont {A.}~\bibnamefont {Luican}}, \bibinfo
  {author} {\bibfnamefont {J.~M.~B.}\ \bibnamefont {Lopes~dos Santos}},
  \bibinfo {author} {\bibfnamefont {A.~H.}\ \bibnamefont {Castro~Neto}},
  \bibinfo {author} {\bibfnamefont {A.}~\bibnamefont {Reina}}, \bibinfo
  {author} {\bibfnamefont {J.}~\bibnamefont {Kong}}, \ and\ \bibinfo {author}
  {\bibfnamefont {E.~Y.}\ \bibnamefont {Andrei}},\ }\bibfield  {title}
  {\enquote {\bibinfo {title} {Observation of van hove singularities in twisted
  graphene layers},}\ }\href {\doibase 10.1038/nphys1463} {\bibfield  {journal}
  {\bibinfo  {journal} {Nature Physics}\ }\textbf {\bibinfo {volume} {6}},\
  \bibinfo {pages} {109--113} (\bibinfo {year} {2010})}\BibitemShut {NoStop}%
\bibitem [{\citenamefont {Bistritzer}\ and\ \citenamefont
  {MacDonald}(2011)}]{Bistritzer:2011}%
  \BibitemOpen
  \bibfield  {author} {\bibinfo {author} {\bibfnamefont {Rafi}\ \bibnamefont
  {Bistritzer}}\ and\ \bibinfo {author} {\bibfnamefont {Allan~H.}\ \bibnamefont
  {MacDonald}},\ }\bibfield  {title} {\enquote {\bibinfo {title} {Moir{\'e}
  bands in twisted double-layer graphene},}\ }\href {\doibase
  10.1073/pnas.1108174108} {\bibfield  {journal} {\bibinfo  {journal}
  {Proceedings of the National Academy of Sciences}\ }\textbf {\bibinfo
  {volume} {108}},\ \bibinfo {pages} {12233--12237} (\bibinfo {year}
  {2011})}\BibitemShut {NoStop}%
\bibitem [{\citenamefont {Lopes~dos Santos}\ \emph {et~al.}(2012)\citenamefont
  {Lopes~dos Santos}, \citenamefont {Peres},\ and\ \citenamefont
  {Castro~Neto}}]{Santos:2012}%
  \BibitemOpen
  \bibfield  {author} {\bibinfo {author} {\bibfnamefont {J.~M.~B.}\
  \bibnamefont {Lopes~dos Santos}}, \bibinfo {author} {\bibfnamefont
  {N.~M.~R.}\ \bibnamefont {Peres}}, \ and\ \bibinfo {author} {\bibfnamefont
  {A.~H.}\ \bibnamefont {Castro~Neto}},\ }\bibfield  {title} {\enquote
  {\bibinfo {title} {Continuum model of the twisted graphene bilayer},}\ }\href
  {\doibase 10.1103/PhysRevB.86.155449} {\bibfield  {journal} {\bibinfo
  {journal} {Phys. Rev. B}\ }\textbf {\bibinfo {volume} {86}},\ \bibinfo
  {pages} {155449} (\bibinfo {year} {2012})}\BibitemShut {NoStop}%
\bibitem [{\citenamefont {Cao}\ \emph {et~al.}(2018{\natexlab{a}})\citenamefont
  {Cao}, \citenamefont {Fatemi}, \citenamefont {Demir}, \citenamefont {Fang},
  \citenamefont {Tomarken}, \citenamefont {Luo}, \citenamefont
  {Sanchez-Yamagishi}, \citenamefont {Watanabe}, \citenamefont {Taniguchi},
  \citenamefont {Kaxiras}, \citenamefont {Ashoori},\ and\ \citenamefont
  {Jarillo-Herrero}}]{Cao:2018b}%
  \BibitemOpen
  \bibfield  {author} {\bibinfo {author} {\bibfnamefont {Yuan}\ \bibnamefont
  {Cao}}, \bibinfo {author} {\bibfnamefont {Valla}\ \bibnamefont {Fatemi}},
  \bibinfo {author} {\bibfnamefont {Ahmet}\ \bibnamefont {Demir}}, \bibinfo
  {author} {\bibfnamefont {Shiang}\ \bibnamefont {Fang}}, \bibinfo {author}
  {\bibfnamefont {Spencer~L.}\ \bibnamefont {Tomarken}}, \bibinfo {author}
  {\bibfnamefont {Jason~Y.}\ \bibnamefont {Luo}}, \bibinfo {author}
  {\bibfnamefont {Javier~D.}\ \bibnamefont {Sanchez-Yamagishi}}, \bibinfo
  {author} {\bibfnamefont {Kenji}\ \bibnamefont {Watanabe}}, \bibinfo {author}
  {\bibfnamefont {Takashi}\ \bibnamefont {Taniguchi}}, \bibinfo {author}
  {\bibfnamefont {Efthimios}\ \bibnamefont {Kaxiras}}, \bibinfo {author}
  {\bibfnamefont {Ray~C.}\ \bibnamefont {Ashoori}}, \ and\ \bibinfo {author}
  {\bibfnamefont {Pablo}\ \bibnamefont {Jarillo-Herrero}},\ }\bibfield  {title}
  {\enquote {\bibinfo {title} {Correlated insulator behaviour at half-filling
  in magic-angle graphene superlattices},}\ }\href {\doibase
  10.1038/nature26154} {\bibfield  {journal} {\bibinfo  {journal} {Nature}\
  }\textbf {\bibinfo {volume} {556}},\ \bibinfo {pages} {80--84} (\bibinfo
  {year} {2018}{\natexlab{a}})}\BibitemShut {NoStop}%
\bibitem [{\citenamefont {Cao}\ \emph {et~al.}(2018{\natexlab{b}})\citenamefont
  {Cao}, \citenamefont {Fatemi}, \citenamefont {Fang}, \citenamefont
  {Watanabe}, \citenamefont {Taniguchi}, \citenamefont {Kaxiras},\ and\
  \citenamefont {Jarillo-Herrero}}]{Cao2018}%
  \BibitemOpen
  \bibfield  {author} {\bibinfo {author} {\bibfnamefont {Yuan}\ \bibnamefont
  {Cao}}, \bibinfo {author} {\bibfnamefont {Valla}\ \bibnamefont {Fatemi}},
  \bibinfo {author} {\bibfnamefont {Shiang}\ \bibnamefont {Fang}}, \bibinfo
  {author} {\bibfnamefont {Kenji}\ \bibnamefont {Watanabe}}, \bibinfo {author}
  {\bibfnamefont {Takashi}\ \bibnamefont {Taniguchi}}, \bibinfo {author}
  {\bibfnamefont {Efthimios}\ \bibnamefont {Kaxiras}}, \ and\ \bibinfo {author}
  {\bibfnamefont {Pablo}\ \bibnamefont {Jarillo-Herrero}},\ }\bibfield  {title}
  {\enquote {\bibinfo {title} {Unconventional superconductivity in magic-angle
  graphene superlattices},}\ }\href {\doibase 10.1038/nature26160} {\bibfield
  {journal} {\bibinfo  {journal} {Nature}\ }\textbf {\bibinfo {volume} {556}},\
  \bibinfo {pages} {43--50} (\bibinfo {year} {2018}{\natexlab{b}})}\BibitemShut
  {NoStop}%
\bibitem [{\citenamefont {Kerelsky}\ \emph {et~al.}(2019)\citenamefont
  {Kerelsky}, \citenamefont {McGilly}, \citenamefont {Kennes}, \citenamefont
  {Xian}, \citenamefont {Yankowitz}, \citenamefont {Chen}, \citenamefont
  {Watanabe}, \citenamefont {Taniguchi}, \citenamefont {Hone}, \citenamefont
  {Dean}, \citenamefont {Rubio},\ and\ \citenamefont
  {Pasupathy}}]{Kerelsky:2019}%
  \BibitemOpen
  \bibfield  {author} {\bibinfo {author} {\bibfnamefont {Alexander}\
  \bibnamefont {Kerelsky}}, \bibinfo {author} {\bibfnamefont {Leo~J.}\
  \bibnamefont {McGilly}}, \bibinfo {author} {\bibfnamefont {Dante~M.}\
  \bibnamefont {Kennes}}, \bibinfo {author} {\bibfnamefont {Lede}\ \bibnamefont
  {Xian}}, \bibinfo {author} {\bibfnamefont {Matthew}\ \bibnamefont
  {Yankowitz}}, \bibinfo {author} {\bibfnamefont {Shaowen}\ \bibnamefont
  {Chen}}, \bibinfo {author} {\bibfnamefont {K.}~\bibnamefont {Watanabe}},
  \bibinfo {author} {\bibfnamefont {T.}~\bibnamefont {Taniguchi}}, \bibinfo
  {author} {\bibfnamefont {James}\ \bibnamefont {Hone}}, \bibinfo {author}
  {\bibfnamefont {Cory}\ \bibnamefont {Dean}}, \bibinfo {author} {\bibfnamefont
  {Angel}\ \bibnamefont {Rubio}}, \ and\ \bibinfo {author} {\bibfnamefont
  {Abhay~N.}\ \bibnamefont {Pasupathy}},\ }\bibfield  {title} {\enquote
  {\bibinfo {title} {Maximized electron interactions at the magic angle in
  twisted bilayer graphene},}\ }\href {\doibase 10.1038/s41586-019-1431-9}
  {\bibfield  {journal} {\bibinfo  {journal} {Nature}\ }\textbf {\bibinfo
  {volume} {572}},\ \bibinfo {pages} {95--100} (\bibinfo {year}
  {2019})}\BibitemShut {NoStop}%
\bibitem [{\citenamefont {Xie}\ \emph {et~al.}(2019)\citenamefont {Xie},
  \citenamefont {Lian}, \citenamefont {J{\"a}ck}, \citenamefont {Liu},
  \citenamefont {Chiu}, \citenamefont {Watanabe}, \citenamefont {Taniguchi},
  \citenamefont {Bernevig},\ and\ \citenamefont {Yazdani}}]{Yonglong:2019}%
  \BibitemOpen
  \bibfield  {author} {\bibinfo {author} {\bibfnamefont {Yonglong}\
  \bibnamefont {Xie}}, \bibinfo {author} {\bibfnamefont {Biao}\ \bibnamefont
  {Lian}}, \bibinfo {author} {\bibfnamefont {Berthold}\ \bibnamefont
  {J{\"a}ck}}, \bibinfo {author} {\bibfnamefont {Xiaomeng}\ \bibnamefont
  {Liu}}, \bibinfo {author} {\bibfnamefont {Cheng-Li}\ \bibnamefont {Chiu}},
  \bibinfo {author} {\bibfnamefont {Kenji}\ \bibnamefont {Watanabe}}, \bibinfo
  {author} {\bibfnamefont {Takashi}\ \bibnamefont {Taniguchi}}, \bibinfo
  {author} {\bibfnamefont {B.~Andrei}\ \bibnamefont {Bernevig}}, \ and\
  \bibinfo {author} {\bibfnamefont {Ali}\ \bibnamefont {Yazdani}},\ }\bibfield
  {title} {\enquote {\bibinfo {title} {Spectroscopic signatures of many-body
  correlations in magic-angle twisted bilayer graphene},}\ }\href {\doibase
  10.1038/s41586-019-1422-x} {\bibfield  {journal} {\bibinfo  {journal}
  {Nature}\ }\textbf {\bibinfo {volume} {572}},\ \bibinfo {pages} {101--105}
  (\bibinfo {year} {2019})}\BibitemShut {NoStop}%
\bibitem [{\citenamefont {Jiang}\ \emph {et~al.}(2019)\citenamefont {Jiang},
  \citenamefont {Lai}, \citenamefont {Watanabe}, \citenamefont {Taniguchi},
  \citenamefont {Haule}, \citenamefont {Mao},\ and\ \citenamefont
  {Andrei}}]{Jiang:2019}%
  \BibitemOpen
  \bibfield  {author} {\bibinfo {author} {\bibfnamefont {Yuhang}\ \bibnamefont
  {Jiang}}, \bibinfo {author} {\bibfnamefont {Xinyuan}\ \bibnamefont {Lai}},
  \bibinfo {author} {\bibfnamefont {Kenji}\ \bibnamefont {Watanabe}}, \bibinfo
  {author} {\bibfnamefont {Takashi}\ \bibnamefont {Taniguchi}}, \bibinfo
  {author} {\bibfnamefont {Kristjan}\ \bibnamefont {Haule}}, \bibinfo {author}
  {\bibfnamefont {Jinhai}\ \bibnamefont {Mao}}, \ and\ \bibinfo {author}
  {\bibfnamefont {Eva~Y.}\ \bibnamefont {Andrei}},\ }\bibfield  {title}
  {\enquote {\bibinfo {title} {Charge order and broken rotational symmetry in
  magic-angle twisted bilayer graphene},}\ }\href {\doibase
  10.1038/s41586-019-1460-4} {\bibfield  {journal} {\bibinfo  {journal}
  {Nature}\ }\textbf {\bibinfo {volume} {573}},\ \bibinfo {pages} {91--95}
  (\bibinfo {year} {2019})}\BibitemShut {NoStop}%
\bibitem [{\citenamefont {Choi}\ \emph {et~al.}(2019)\citenamefont {Choi},
  \citenamefont {Kemmer}, \citenamefont {Peng}, \citenamefont {Thomson},
  \citenamefont {Arora}, \citenamefont {Polski}, \citenamefont {Zhang},
  \citenamefont {Ren}, \citenamefont {Alicea}, \citenamefont {Refael},
  \citenamefont {von Oppen}, \citenamefont {Watanabe}, \citenamefont
  {Taniguchi},\ and\ \citenamefont {Nadj-Perge}}]{Choi:2019}%
  \BibitemOpen
  \bibfield  {author} {\bibinfo {author} {\bibfnamefont {Youngjoon}\
  \bibnamefont {Choi}}, \bibinfo {author} {\bibfnamefont {Jeannette}\
  \bibnamefont {Kemmer}}, \bibinfo {author} {\bibfnamefont {Yang}\ \bibnamefont
  {Peng}}, \bibinfo {author} {\bibfnamefont {Alex}\ \bibnamefont {Thomson}},
  \bibinfo {author} {\bibfnamefont {Harpreet}\ \bibnamefont {Arora}}, \bibinfo
  {author} {\bibfnamefont {Robert}\ \bibnamefont {Polski}}, \bibinfo {author}
  {\bibfnamefont {Yiran}\ \bibnamefont {Zhang}}, \bibinfo {author}
  {\bibfnamefont {Hechen}\ \bibnamefont {Ren}}, \bibinfo {author}
  {\bibfnamefont {Jason}\ \bibnamefont {Alicea}}, \bibinfo {author}
  {\bibfnamefont {Gil}\ \bibnamefont {Refael}}, \bibinfo {author}
  {\bibfnamefont {Felix}\ \bibnamefont {von Oppen}}, \bibinfo {author}
  {\bibfnamefont {Kenji}\ \bibnamefont {Watanabe}}, \bibinfo {author}
  {\bibfnamefont {Takashi}\ \bibnamefont {Taniguchi}}, \ and\ \bibinfo {author}
  {\bibfnamefont {Stevan}\ \bibnamefont {Nadj-Perge}},\ }\bibfield  {title}
  {\enquote {\bibinfo {title} {Electronic correlations in twisted bilayer
  graphene near the magic angle},}\ }\href {\doibase 10.1038/s41567-019-0606-5}
  {\bibfield  {journal} {\bibinfo  {journal} {Nature Physics}\ }\textbf
  {\bibinfo {volume} {15}},\ \bibinfo {pages} {1174--1180} (\bibinfo {year}
  {2019})}\BibitemShut {NoStop}%
\bibitem [{\citenamefont {Yankowitz}\ \emph {et~al.}(2019)\citenamefont
  {Yankowitz}, \citenamefont {Chen}, \citenamefont {Polshyn}, \citenamefont
  {Zhang}, \citenamefont {Watanabe}, \citenamefont {Taniguchi}, \citenamefont
  {Graf}, \citenamefont {Young},\ and\ \citenamefont {Dean}}]{Yankowitz:2019}%
  \BibitemOpen
  \bibfield  {author} {\bibinfo {author} {\bibfnamefont {Matthew}\ \bibnamefont
  {Yankowitz}}, \bibinfo {author} {\bibfnamefont {Shaowen}\ \bibnamefont
  {Chen}}, \bibinfo {author} {\bibfnamefont {Hryhoriy}\ \bibnamefont
  {Polshyn}}, \bibinfo {author} {\bibfnamefont {Yuxuan}\ \bibnamefont {Zhang}},
  \bibinfo {author} {\bibfnamefont {K.}~\bibnamefont {Watanabe}}, \bibinfo
  {author} {\bibfnamefont {T.}~\bibnamefont {Taniguchi}}, \bibinfo {author}
  {\bibfnamefont {David}\ \bibnamefont {Graf}}, \bibinfo {author}
  {\bibfnamefont {Andrea~F.}\ \bibnamefont {Young}}, \ and\ \bibinfo {author}
  {\bibfnamefont {Cory~R.}\ \bibnamefont {Dean}},\ }\bibfield  {title}
  {\enquote {\bibinfo {title} {Tuning superconductivity in twisted bilayer
  graphene},}\ }\href {\doibase 10.1126/science.aav1910} {\bibfield  {journal}
  {\bibinfo  {journal} {Science}\ }\textbf {\bibinfo {volume} {363}},\ \bibinfo
  {pages} {1059--1064} (\bibinfo {year} {2019})}\BibitemShut {NoStop}%
\bibitem [{\citenamefont {Lu}\ \emph {et~al.}(2019)\citenamefont {Lu},
  \citenamefont {Stepanov}, \citenamefont {Yang}, \citenamefont {Xie},
  \citenamefont {Aamir}, \citenamefont {Das}, \citenamefont {Urgell},
  \citenamefont {Watanabe}, \citenamefont {Taniguchi}, \citenamefont {Zhang},
  \citenamefont {Bachtold}, \citenamefont {MacDonald},\ and\ \citenamefont
  {Efetov}}]{Xiaobo:2019}%
  \BibitemOpen
  \bibfield  {author} {\bibinfo {author} {\bibfnamefont {Xiaobo}\ \bibnamefont
  {Lu}}, \bibinfo {author} {\bibfnamefont {Petr}\ \bibnamefont {Stepanov}},
  \bibinfo {author} {\bibfnamefont {Wei}\ \bibnamefont {Yang}}, \bibinfo
  {author} {\bibfnamefont {Ming}\ \bibnamefont {Xie}}, \bibinfo {author}
  {\bibfnamefont {Mohammed~Ali}\ \bibnamefont {Aamir}}, \bibinfo {author}
  {\bibfnamefont {Ipsita}\ \bibnamefont {Das}}, \bibinfo {author}
  {\bibfnamefont {Carles}\ \bibnamefont {Urgell}}, \bibinfo {author}
  {\bibfnamefont {Kenji}\ \bibnamefont {Watanabe}}, \bibinfo {author}
  {\bibfnamefont {Takashi}\ \bibnamefont {Taniguchi}}, \bibinfo {author}
  {\bibfnamefont {Guangyu}\ \bibnamefont {Zhang}}, \bibinfo {author}
  {\bibfnamefont {Adrian}\ \bibnamefont {Bachtold}}, \bibinfo {author}
  {\bibfnamefont {Allan~H.}\ \bibnamefont {MacDonald}}, \ and\ \bibinfo
  {author} {\bibfnamefont {Dmitri~K.}\ \bibnamefont {Efetov}},\ }\bibfield
  {title} {\enquote {\bibinfo {title} {Superconductors, orbital magnets and
  correlated states in magic-angle bilayer graphene},}\ }\href {\doibase
  10.1038/s41586-019-1695-0} {\bibfield  {journal} {\bibinfo  {journal}
  {Nature}\ }\textbf {\bibinfo {volume} {574}},\ \bibinfo {pages} {653--657}
  (\bibinfo {year} {2019})}\BibitemShut {NoStop}%
\bibitem [{\citenamefont {Andrei}\ and\ \citenamefont
  {MacDonald}(2020)}]{Andrei:2020}%
  \BibitemOpen
  \bibfield  {author} {\bibinfo {author} {\bibfnamefont {Eva~Y.}\ \bibnamefont
  {Andrei}}\ and\ \bibinfo {author} {\bibfnamefont {Allan~H.}\ \bibnamefont
  {MacDonald}},\ }\bibfield  {title} {\enquote {\bibinfo {title} {Graphene
  bilayers with a twist},}\ }\href {\doibase 10.1038/s41563-020-00840-0}
  {\bibfield  {journal} {\bibinfo  {journal} {Nature Materials}\ }\textbf
  {\bibinfo {volume} {19}},\ \bibinfo {pages} {1265--1275} (\bibinfo {year}
  {2020})}\BibitemShut {NoStop}%
\bibitem [{\citenamefont {Kennes}\ \emph {et~al.}(2021)\citenamefont {Kennes},
  \citenamefont {Claassen}, \citenamefont {Xian}, \citenamefont {Georges},
  \citenamefont {Millis}, \citenamefont {Hone}, \citenamefont {Dean},
  \citenamefont {Basov}, \citenamefont {Pasupathy},\ and\ \citenamefont
  {Rubio}}]{Kennes:2021}%
  \BibitemOpen
  \bibfield  {author} {\bibinfo {author} {\bibfnamefont {Dante~M.}\
  \bibnamefont {Kennes}}, \bibinfo {author} {\bibfnamefont {Martin}\
  \bibnamefont {Claassen}}, \bibinfo {author} {\bibfnamefont {Lede}\
  \bibnamefont {Xian}}, \bibinfo {author} {\bibfnamefont {Antoine}\
  \bibnamefont {Georges}}, \bibinfo {author} {\bibfnamefont {Andrew~J.}\
  \bibnamefont {Millis}}, \bibinfo {author} {\bibfnamefont {James}\
  \bibnamefont {Hone}}, \bibinfo {author} {\bibfnamefont {Cory~R.}\
  \bibnamefont {Dean}}, \bibinfo {author} {\bibfnamefont {D.~N.}\ \bibnamefont
  {Basov}}, \bibinfo {author} {\bibfnamefont {Abhay~N.}\ \bibnamefont
  {Pasupathy}}, \ and\ \bibinfo {author} {\bibfnamefont {Angel}\ \bibnamefont
  {Rubio}},\ }\bibfield  {title} {\enquote {\bibinfo {title} {Moir{\'e}
  heterostructures as a condensed-matter quantum simulator},}\ }\href {\doibase
  10.1038/s41567-020-01154-3} {\bibfield  {journal} {\bibinfo  {journal}
  {Nature Physics}\ }\textbf {\bibinfo {volume} {17}},\ \bibinfo {pages}
  {155--163} (\bibinfo {year} {2021})}\BibitemShut {NoStop}%
\bibitem [{\citenamefont {Damascelli}\ \emph {et~al.}(2003)\citenamefont
  {Damascelli}, \citenamefont {Hussain},\ and\ \citenamefont
  {Shen}}]{Damascelli:2003}%
  \BibitemOpen
  \bibfield  {author} {\bibinfo {author} {\bibfnamefont {Andrea}\ \bibnamefont
  {Damascelli}}, \bibinfo {author} {\bibfnamefont {Zahid}\ \bibnamefont
  {Hussain}}, \ and\ \bibinfo {author} {\bibfnamefont {Zhi-Xun}\ \bibnamefont
  {Shen}},\ }\bibfield  {title} {\enquote {\bibinfo {title} {Angle-resolved
  photoemission studies of the cuprate superconductors},}\ }\href {\doibase
  10.1103/RevModPhys.75.473} {\bibfield  {journal} {\bibinfo  {journal} {Rev.
  Mod. Phys.}\ }\textbf {\bibinfo {volume} {75}},\ \bibinfo {pages} {473--541}
  (\bibinfo {year} {2003})}\BibitemShut {NoStop}%
\bibitem [{\citenamefont {Sobota}\ \emph {et~al.}(2021)\citenamefont {Sobota},
  \citenamefont {He},\ and\ \citenamefont {Shen}}]{Sobota:2021}%
  \BibitemOpen
  \bibfield  {author} {\bibinfo {author} {\bibfnamefont {Jonathan~A.}\
  \bibnamefont {Sobota}}, \bibinfo {author} {\bibfnamefont {Yu}~\bibnamefont
  {He}}, \ and\ \bibinfo {author} {\bibfnamefont {Zhi-Xun}\ \bibnamefont
  {Shen}},\ }\bibfield  {title} {\enquote {\bibinfo {title} {Angle-resolved
  photoemission studies of quantum materials},}\ }\href {\doibase
  10.1103/RevModPhys.93.025006} {\bibfield  {journal} {\bibinfo  {journal}
  {Rev. Mod. Phys.}\ }\textbf {\bibinfo {volume} {93}},\ \bibinfo {pages}
  {025006} (\bibinfo {year} {2021})}\BibitemShut {NoStop}%
\bibitem [{\citenamefont {Utama}\ \emph {et~al.}(2021)\citenamefont {Utama},
  \citenamefont {Koch}, \citenamefont {Lee}, \citenamefont {Leconte},
  \citenamefont {Li}, \citenamefont {Zhao}, \citenamefont {Jiang},
  \citenamefont {Zhu}, \citenamefont {Watanabe}, \citenamefont {Taniguchi},
  \citenamefont {Ashby}, \citenamefont {Weber-Bargioni}, \citenamefont {Zettl},
  \citenamefont {Jozwiak}, \citenamefont {Jung}, \citenamefont {Rotenberg},
  \citenamefont {Bostwick},\ and\ \citenamefont {Wang}}]{Utama:2021}%
  \BibitemOpen
  \bibfield  {author} {\bibinfo {author} {\bibfnamefont {M.~Iqbal~Bakti}\
  \bibnamefont {Utama}}, \bibinfo {author} {\bibfnamefont {Roland~J.}\
  \bibnamefont {Koch}}, \bibinfo {author} {\bibfnamefont {Kyunghoon}\
  \bibnamefont {Lee}}, \bibinfo {author} {\bibfnamefont {Nicolas}\ \bibnamefont
  {Leconte}}, \bibinfo {author} {\bibfnamefont {Hongyuan}\ \bibnamefont {Li}},
  \bibinfo {author} {\bibfnamefont {Sihan}\ \bibnamefont {Zhao}}, \bibinfo
  {author} {\bibfnamefont {Lili}\ \bibnamefont {Jiang}}, \bibinfo {author}
  {\bibfnamefont {Jiayi}\ \bibnamefont {Zhu}}, \bibinfo {author} {\bibfnamefont
  {Kenji}\ \bibnamefont {Watanabe}}, \bibinfo {author} {\bibfnamefont
  {Takashi}\ \bibnamefont {Taniguchi}}, \bibinfo {author} {\bibfnamefont
  {Paul~D.}\ \bibnamefont {Ashby}}, \bibinfo {author} {\bibfnamefont
  {Alexander}\ \bibnamefont {Weber-Bargioni}}, \bibinfo {author} {\bibfnamefont
  {Alex}\ \bibnamefont {Zettl}}, \bibinfo {author} {\bibfnamefont {Chris}\
  \bibnamefont {Jozwiak}}, \bibinfo {author} {\bibfnamefont {Jeil}\
  \bibnamefont {Jung}}, \bibinfo {author} {\bibfnamefont {Eli}\ \bibnamefont
  {Rotenberg}}, \bibinfo {author} {\bibfnamefont {Aaron}\ \bibnamefont
  {Bostwick}}, \ and\ \bibinfo {author} {\bibfnamefont {Feng}\ \bibnamefont
  {Wang}},\ }\bibfield  {title} {\enquote {\bibinfo {title} {Visualization of
  the flat electronic band in twisted bilayer graphene near the magic angle
  twist},}\ }\href {\doibase 10.1038/s41567-020-0974-x} {\bibfield  {journal}
  {\bibinfo  {journal} {Nature Physics}\ }\textbf {\bibinfo {volume} {17}},\
  \bibinfo {pages} {184--188} (\bibinfo {year} {2021})}\BibitemShut {NoStop}%
\bibitem [{\citenamefont {Lisi}\ \emph {et~al.}(2021)\citenamefont {Lisi},
  \citenamefont {Lu}, \citenamefont {Benschop}, \citenamefont {de~Jong},
  \citenamefont {Stepanov}, \citenamefont {Duran}, \citenamefont {Margot},
  \citenamefont {Cucchi}, \citenamefont {Cappelli}, \citenamefont {Hunter},
  \citenamefont {Tamai}, \citenamefont {Kandyba}, \citenamefont {Giampietri},
  \citenamefont {Barinov}, \citenamefont {Jobst}, \citenamefont {Stalman},
  \citenamefont {Leeuwenhoek}, \citenamefont {Watanabe}, \citenamefont
  {Taniguchi}, \citenamefont {Rademaker}, \citenamefont {van~der Molen},
  \citenamefont {Allan}, \citenamefont {Efetov},\ and\ \citenamefont
  {Baumberger}}]{Lisi:2021}%
  \BibitemOpen
  \bibfield  {author} {\bibinfo {author} {\bibfnamefont {Simone}\ \bibnamefont
  {Lisi}}, \bibinfo {author} {\bibfnamefont {Xiaobo}\ \bibnamefont {Lu}},
  \bibinfo {author} {\bibfnamefont {Tjerk}\ \bibnamefont {Benschop}}, \bibinfo
  {author} {\bibfnamefont {Tobias~A.}\ \bibnamefont {de~Jong}}, \bibinfo
  {author} {\bibfnamefont {Petr}\ \bibnamefont {Stepanov}}, \bibinfo {author}
  {\bibfnamefont {Jose~R.}\ \bibnamefont {Duran}}, \bibinfo {author}
  {\bibfnamefont {Florian}\ \bibnamefont {Margot}}, \bibinfo {author}
  {\bibfnamefont {Ir{\`e}ne}\ \bibnamefont {Cucchi}}, \bibinfo {author}
  {\bibfnamefont {Edoardo}\ \bibnamefont {Cappelli}}, \bibinfo {author}
  {\bibfnamefont {Andrew}\ \bibnamefont {Hunter}}, \bibinfo {author}
  {\bibfnamefont {Anna}\ \bibnamefont {Tamai}}, \bibinfo {author}
  {\bibfnamefont {Viktor}\ \bibnamefont {Kandyba}}, \bibinfo {author}
  {\bibfnamefont {Alessio}\ \bibnamefont {Giampietri}}, \bibinfo {author}
  {\bibfnamefont {Alexei}\ \bibnamefont {Barinov}}, \bibinfo {author}
  {\bibfnamefont {Johannes}\ \bibnamefont {Jobst}}, \bibinfo {author}
  {\bibfnamefont {Vincent}\ \bibnamefont {Stalman}}, \bibinfo {author}
  {\bibfnamefont {Maarten}\ \bibnamefont {Leeuwenhoek}}, \bibinfo {author}
  {\bibfnamefont {Kenji}\ \bibnamefont {Watanabe}}, \bibinfo {author}
  {\bibfnamefont {Takashi}\ \bibnamefont {Taniguchi}}, \bibinfo {author}
  {\bibfnamefont {Louk}\ \bibnamefont {Rademaker}}, \bibinfo {author}
  {\bibfnamefont {Sense~Jan}\ \bibnamefont {van~der Molen}}, \bibinfo {author}
  {\bibfnamefont {Milan~P.}\ \bibnamefont {Allan}}, \bibinfo {author}
  {\bibfnamefont {Dmitri~K.}\ \bibnamefont {Efetov}}, \ and\ \bibinfo {author}
  {\bibfnamefont {Felix}\ \bibnamefont {Baumberger}},\ }\bibfield  {title}
  {\enquote {\bibinfo {title} {Observation of flat bands in twisted bilayer
  graphene},}\ }\href {\doibase 10.1038/s41567-020-01041-x} {\bibfield
  {journal} {\bibinfo  {journal} {Nature Physics}\ }\textbf {\bibinfo {volume}
  {17}},\ \bibinfo {pages} {189--193} (\bibinfo {year} {2021})}\BibitemShut
  {NoStop}%
\bibitem [{\citenamefont {Li}\ \emph {et~al.}(2022)\citenamefont {Li},
  \citenamefont {Zhang}, \citenamefont {Chen}, \citenamefont {Wei},
  \citenamefont {Zhang}, \citenamefont {Xiao}, \citenamefont {Gao},
  \citenamefont {Chen}, \citenamefont {Liang}, \citenamefont {Pei},
  \citenamefont {Xu}, \citenamefont {Watanabe}, \citenamefont {Taniguchi},
  \citenamefont {Yang}, \citenamefont {Miao}, \citenamefont {Liu},
  \citenamefont {Cheng}, \citenamefont {Wang}, \citenamefont {Chen},\ and\
  \citenamefont {Liu}}]{Yiwei:2022}%
  \BibitemOpen
  \bibfield  {author} {\bibinfo {author} {\bibfnamefont {Yiwei}\ \bibnamefont
  {Li}}, \bibinfo {author} {\bibfnamefont {Shihao}\ \bibnamefont {Zhang}},
  \bibinfo {author} {\bibfnamefont {Fanqiang}\ \bibnamefont {Chen}}, \bibinfo
  {author} {\bibfnamefont {Liyang}\ \bibnamefont {Wei}}, \bibinfo {author}
  {\bibfnamefont {Zonglin}\ \bibnamefont {Zhang}}, \bibinfo {author}
  {\bibfnamefont {Hanbo}\ \bibnamefont {Xiao}}, \bibinfo {author}
  {\bibfnamefont {Han}\ \bibnamefont {Gao}}, \bibinfo {author} {\bibfnamefont
  {Moyu}\ \bibnamefont {Chen}}, \bibinfo {author} {\bibfnamefont {Shijun}\
  \bibnamefont {Liang}}, \bibinfo {author} {\bibfnamefont {Ding}\ \bibnamefont
  {Pei}}, \bibinfo {author} {\bibfnamefont {Lixuan}\ \bibnamefont {Xu}},
  \bibinfo {author} {\bibfnamefont {Kenji}\ \bibnamefont {Watanabe}}, \bibinfo
  {author} {\bibfnamefont {Takashi}\ \bibnamefont {Taniguchi}}, \bibinfo
  {author} {\bibfnamefont {Lexian}\ \bibnamefont {Yang}}, \bibinfo {author}
  {\bibfnamefont {Feng}\ \bibnamefont {Miao}}, \bibinfo {author} {\bibfnamefont
  {Jianpeng}\ \bibnamefont {Liu}}, \bibinfo {author} {\bibfnamefont {Bin}\
  \bibnamefont {Cheng}}, \bibinfo {author} {\bibfnamefont {Meixiao}\
  \bibnamefont {Wang}}, \bibinfo {author} {\bibfnamefont {Yulin}\ \bibnamefont
  {Chen}}, \ and\ \bibinfo {author} {\bibfnamefont {Zhongkai}\ \bibnamefont
  {Liu}},\ }\bibfield  {title} {\enquote {\bibinfo {title} {Observation of
  coexisting dirac bands and moiré flat bands in magic-angle twisted trilayer
  graphene},}\ }\href {\doibase https://doi.org/10.1002/adma.202205996}
  {\bibfield  {journal} {\bibinfo  {journal} {Advanced Materials}\ }\textbf
  {\bibinfo {volume} {34}},\ \bibinfo {pages} {2205996} (\bibinfo {year}
  {2022})}\BibitemShut {NoStop}%
\bibitem [{\citenamefont {Kim}\ \emph {et~al.}(2016)\citenamefont {Kim},
  \citenamefont {Yankowitz}, \citenamefont {Fallahazad}, \citenamefont {Kang},
  \citenamefont {Movva}, \citenamefont {Huang}, \citenamefont {Larentis},
  \citenamefont {Corbet}, \citenamefont {Taniguchi}, \citenamefont {Watanabe},
  \citenamefont {Banerjee}, \citenamefont {LeRoy},\ and\ \citenamefont
  {Tutuc}}]{Kyounghwan:2016}%
  \BibitemOpen
  \bibfield  {author} {\bibinfo {author} {\bibfnamefont {Kyounghwan}\
  \bibnamefont {Kim}}, \bibinfo {author} {\bibfnamefont {Matthew}\ \bibnamefont
  {Yankowitz}}, \bibinfo {author} {\bibfnamefont {Babak}\ \bibnamefont
  {Fallahazad}}, \bibinfo {author} {\bibfnamefont {Sangwoo}\ \bibnamefont
  {Kang}}, \bibinfo {author} {\bibfnamefont {Hema C.~P.}\ \bibnamefont
  {Movva}}, \bibinfo {author} {\bibfnamefont {Shengqiang}\ \bibnamefont
  {Huang}}, \bibinfo {author} {\bibfnamefont {Stefano}\ \bibnamefont
  {Larentis}}, \bibinfo {author} {\bibfnamefont {Chris~M.}\ \bibnamefont
  {Corbet}}, \bibinfo {author} {\bibfnamefont {Takashi}\ \bibnamefont
  {Taniguchi}}, \bibinfo {author} {\bibfnamefont {Kenji}\ \bibnamefont
  {Watanabe}}, \bibinfo {author} {\bibfnamefont {Sanjay~K.}\ \bibnamefont
  {Banerjee}}, \bibinfo {author} {\bibfnamefont {Brian~J.}\ \bibnamefont
  {LeRoy}}, \ and\ \bibinfo {author} {\bibfnamefont {Emanuel}\ \bibnamefont
  {Tutuc}},\ }\bibfield  {title} {\enquote {\bibinfo {title} {van der waals
  heterostructures with high accuracy rotational alignment},}\ }\href {\doibase
  10.1021/acs.nanolett.5b05263} {\bibfield  {journal} {\bibinfo  {journal}
  {Nano Letters}\ }\textbf {\bibinfo {volume} {16}},\ \bibinfo {pages}
  {1989--1995} (\bibinfo {year} {2016})}\BibitemShut {NoStop}%
\bibitem [{\citenamefont {Kim}\ \emph {et~al.}(2017)\citenamefont {Kim},
  \citenamefont {DaSilva}, \citenamefont {Huang}, \citenamefont {Fallahazad},
  \citenamefont {Larentis}, \citenamefont {Taniguchi}, \citenamefont
  {Watanabe}, \citenamefont {LeRoy}, \citenamefont {MacDonald},\ and\
  \citenamefont {Tutuc}}]{Kyounghwan:2017}%
  \BibitemOpen
  \bibfield  {author} {\bibinfo {author} {\bibfnamefont {Kyounghwan}\
  \bibnamefont {Kim}}, \bibinfo {author} {\bibfnamefont {Ashley}\ \bibnamefont
  {DaSilva}}, \bibinfo {author} {\bibfnamefont {Shengqiang}\ \bibnamefont
  {Huang}}, \bibinfo {author} {\bibfnamefont {Babak}\ \bibnamefont
  {Fallahazad}}, \bibinfo {author} {\bibfnamefont {Stefano}\ \bibnamefont
  {Larentis}}, \bibinfo {author} {\bibfnamefont {Takashi}\ \bibnamefont
  {Taniguchi}}, \bibinfo {author} {\bibfnamefont {Kenji}\ \bibnamefont
  {Watanabe}}, \bibinfo {author} {\bibfnamefont {Brian~J.}\ \bibnamefont
  {LeRoy}}, \bibinfo {author} {\bibfnamefont {Allan~H.}\ \bibnamefont
  {MacDonald}}, \ and\ \bibinfo {author} {\bibfnamefont {Emanuel}\ \bibnamefont
  {Tutuc}},\ }\bibfield  {title} {\enquote {\bibinfo {title} {Tunable moiré
  bands and strong correlations in small-twist-angle bilayer graphene},}\
  }\href {\doibase 10.1073/pnas.1620140114} {\bibfield  {journal} {\bibinfo
  {journal} {Proceedings of the National Academy of Sciences}\ }\textbf
  {\bibinfo {volume} {114}},\ \bibinfo {pages} {3364--3369} (\bibinfo {year}
  {2017})}\BibitemShut {NoStop}%
\bibitem [{\citenamefont {Uri}\ \emph {et~al.}(2020)\citenamefont {Uri},
  \citenamefont {Grover}, \citenamefont {Cao}, \citenamefont {Crosse},
  \citenamefont {Bagani}, \citenamefont {Rodan-Legrain}, \citenamefont
  {Myasoedov}, \citenamefont {Watanabe}, \citenamefont {Taniguchi},
  \citenamefont {Moon}, \citenamefont {Koshino}, \citenamefont
  {Jarillo-Herrero},\ and\ \citenamefont {Zeldov}}]{Uri:2020}%
  \BibitemOpen
  \bibfield  {author} {\bibinfo {author} {\bibfnamefont {A.}~\bibnamefont
  {Uri}}, \bibinfo {author} {\bibfnamefont {S.}~\bibnamefont {Grover}},
  \bibinfo {author} {\bibfnamefont {Y.}~\bibnamefont {Cao}}, \bibinfo {author}
  {\bibfnamefont {J.~A.}\ \bibnamefont {Crosse}}, \bibinfo {author}
  {\bibfnamefont {K.}~\bibnamefont {Bagani}}, \bibinfo {author} {\bibfnamefont
  {D.}~\bibnamefont {Rodan-Legrain}}, \bibinfo {author} {\bibfnamefont
  {Y.}~\bibnamefont {Myasoedov}}, \bibinfo {author} {\bibfnamefont
  {K.}~\bibnamefont {Watanabe}}, \bibinfo {author} {\bibfnamefont
  {T.}~\bibnamefont {Taniguchi}}, \bibinfo {author} {\bibfnamefont
  {P.}~\bibnamefont {Moon}}, \bibinfo {author} {\bibfnamefont {M.}~\bibnamefont
  {Koshino}}, \bibinfo {author} {\bibfnamefont {P.}~\bibnamefont
  {Jarillo-Herrero}}, \ and\ \bibinfo {author} {\bibfnamefont {E.}~\bibnamefont
  {Zeldov}},\ }\bibfield  {title} {\enquote {\bibinfo {title} {Mapping the
  twist-angle disorder and landau levels in magic-angle graphene},}\ }\href
  {\doibase 10.1038/s41586-020-2255-3} {\bibfield  {journal} {\bibinfo
  {journal} {Nature}\ }\textbf {\bibinfo {volume} {581}},\ \bibinfo {pages}
  {47--52} (\bibinfo {year} {2020})}\BibitemShut {NoStop}%
\bibitem [{\citenamefont {Lau}\ \emph {et~al.}(2022)\citenamefont {Lau},
  \citenamefont {Bockrath}, \citenamefont {Mak},\ and\ \citenamefont
  {Zhang}}]{Lau:2022}%
  \BibitemOpen
  \bibfield  {author} {\bibinfo {author} {\bibfnamefont {Chun~Ning}\
  \bibnamefont {Lau}}, \bibinfo {author} {\bibfnamefont {Marc~W.}\ \bibnamefont
  {Bockrath}}, \bibinfo {author} {\bibfnamefont {Kin~Fai}\ \bibnamefont {Mak}},
  \ and\ \bibinfo {author} {\bibfnamefont {Fan}\ \bibnamefont {Zhang}},\
  }\bibfield  {title} {\enquote {\bibinfo {title} {Reproducibility in the
  fabrication and physics of moir{\'e} materials},}\ }\href {\doibase
  10.1038/s41586-021-04173-z} {\bibfield  {journal} {\bibinfo  {journal}
  {Nature}\ }\textbf {\bibinfo {volume} {602}},\ \bibinfo {pages} {41--50}
  (\bibinfo {year} {2022})}\BibitemShut {NoStop}%
\bibitem [{\citenamefont {Joucken}\ \emph {et~al.}(2019)\citenamefont
  {Joucken}, \citenamefont {Avila}, \citenamefont {Ge}, \citenamefont
  {Quezada-Lopez}, \citenamefont {Yi}, \citenamefont {Le~Goff}, \citenamefont
  {Baudin}, \citenamefont {Davenport}, \citenamefont {Watanabe}, \citenamefont
  {Taniguchi}, \citenamefont {Asensio},\ and\ \citenamefont
  {Velasco}}]{Joucken:2019}%
  \BibitemOpen
  \bibfield  {author} {\bibinfo {author} {\bibfnamefont {Fr{\'e}d{\'e}ric}\
  \bibnamefont {Joucken}}, \bibinfo {author} {\bibfnamefont {Jose}\
  \bibnamefont {Avila}}, \bibinfo {author} {\bibfnamefont {Zhehao}\
  \bibnamefont {Ge}}, \bibinfo {author} {\bibfnamefont {Eberth~A.}\
  \bibnamefont {Quezada-Lopez}}, \bibinfo {author} {\bibfnamefont {Hemian}\
  \bibnamefont {Yi}}, \bibinfo {author} {\bibfnamefont {Romaric}\ \bibnamefont
  {Le~Goff}}, \bibinfo {author} {\bibfnamefont {Emmanuel}\ \bibnamefont
  {Baudin}}, \bibinfo {author} {\bibfnamefont {John~L.}\ \bibnamefont
  {Davenport}}, \bibinfo {author} {\bibfnamefont {Kenji}\ \bibnamefont
  {Watanabe}}, \bibinfo {author} {\bibfnamefont {Takashi}\ \bibnamefont
  {Taniguchi}}, \bibinfo {author} {\bibfnamefont {Maria~Carmen}\ \bibnamefont
  {Asensio}}, \ and\ \bibinfo {author} {\bibfnamefont {Jairo}\ \bibnamefont
  {Velasco}},\ }\bibfield  {title} {\enquote {\bibinfo {title} {Visualizing the
  effect of an electrostatic gate with angle-resolved photoemission
  spectroscopy},}\ }\href {\doibase 10.1021/acs.nanolett.9b00649} {\bibfield
  {journal} {\bibinfo  {journal} {Nano Letters}\ }\textbf {\bibinfo {volume}
  {19}},\ \bibinfo {pages} {2682--2687} (\bibinfo {year} {2019})}\BibitemShut
  {NoStop}%
\bibitem [{\citenamefont {Nguyen}\ \emph {et~al.}(2019)\citenamefont {Nguyen},
  \citenamefont {Teutsch}, \citenamefont {Wilson}, \citenamefont {Kahn},
  \citenamefont {Xia}, \citenamefont {Graham}, \citenamefont {Kandyba},
  \citenamefont {Giampietri}, \citenamefont {Barinov}, \citenamefont
  {Constantinescu}, \citenamefont {Yeung}, \citenamefont {Hine}, \citenamefont
  {Xu}, \citenamefont {Cobden},\ and\ \citenamefont {Wilson}}]{Nguyen:2019}%
  \BibitemOpen
  \bibfield  {author} {\bibinfo {author} {\bibfnamefont {Paul~V.}\ \bibnamefont
  {Nguyen}}, \bibinfo {author} {\bibfnamefont {Natalie~C.}\ \bibnamefont
  {Teutsch}}, \bibinfo {author} {\bibfnamefont {Nathan~P.}\ \bibnamefont
  {Wilson}}, \bibinfo {author} {\bibfnamefont {Joshua}\ \bibnamefont {Kahn}},
  \bibinfo {author} {\bibfnamefont {Xue}\ \bibnamefont {Xia}}, \bibinfo
  {author} {\bibfnamefont {Abigail~J.}\ \bibnamefont {Graham}}, \bibinfo
  {author} {\bibfnamefont {Viktor}\ \bibnamefont {Kandyba}}, \bibinfo {author}
  {\bibfnamefont {Alessio}\ \bibnamefont {Giampietri}}, \bibinfo {author}
  {\bibfnamefont {Alexei}\ \bibnamefont {Barinov}}, \bibinfo {author}
  {\bibfnamefont {Gabriel~C.}\ \bibnamefont {Constantinescu}}, \bibinfo
  {author} {\bibfnamefont {Nelson}\ \bibnamefont {Yeung}}, \bibinfo {author}
  {\bibfnamefont {Nicholas D.~M.}\ \bibnamefont {Hine}}, \bibinfo {author}
  {\bibfnamefont {Xiaodong}\ \bibnamefont {Xu}}, \bibinfo {author}
  {\bibfnamefont {David~H.}\ \bibnamefont {Cobden}}, \ and\ \bibinfo {author}
  {\bibfnamefont {Neil~R.}\ \bibnamefont {Wilson}},\ }\bibfield  {title}
  {\enquote {\bibinfo {title} {Visualizing electrostatic gating effects in
  two-dimensional heterostructures},}\ }\href {\doibase
  10.1038/s41586-019-1402-1} {\bibfield  {journal} {\bibinfo  {journal}
  {Nature}\ }\textbf {\bibinfo {volume} {572}},\ \bibinfo {pages} {220--223}
  (\bibinfo {year} {2019})}\BibitemShut {NoStop}%
\bibitem [{\citenamefont {Muzzio}\ \emph {et~al.}(2020)\citenamefont {Muzzio},
  \citenamefont {Jones}, \citenamefont {Curcio}, \citenamefont {Biswas},
  \citenamefont {Miwa}, \citenamefont {Hofmann}, \citenamefont {Watanabe},
  \citenamefont {Taniguchi}, \citenamefont {Singh}, \citenamefont {Jozwiak},
  \citenamefont {Rotenberg}, \citenamefont {Bostwick}, \citenamefont {Koch},
  \citenamefont {Ulstrup},\ and\ \citenamefont {Katoch}}]{muzzio2020}%
  \BibitemOpen
  \bibfield  {author} {\bibinfo {author} {\bibfnamefont {Ryan}\ \bibnamefont
  {Muzzio}}, \bibinfo {author} {\bibfnamefont {Alfred J.~H.}\ \bibnamefont
  {Jones}}, \bibinfo {author} {\bibfnamefont {Davide}\ \bibnamefont {Curcio}},
  \bibinfo {author} {\bibfnamefont {Deepnarayan}\ \bibnamefont {Biswas}},
  \bibinfo {author} {\bibfnamefont {Jill~A.}\ \bibnamefont {Miwa}}, \bibinfo
  {author} {\bibfnamefont {Philip}\ \bibnamefont {Hofmann}}, \bibinfo {author}
  {\bibfnamefont {Kenji}\ \bibnamefont {Watanabe}}, \bibinfo {author}
  {\bibfnamefont {Takashi}\ \bibnamefont {Taniguchi}}, \bibinfo {author}
  {\bibfnamefont {Simranjeet}\ \bibnamefont {Singh}}, \bibinfo {author}
  {\bibfnamefont {Chris}\ \bibnamefont {Jozwiak}}, \bibinfo {author}
  {\bibfnamefont {Eli}\ \bibnamefont {Rotenberg}}, \bibinfo {author}
  {\bibfnamefont {Aaron}\ \bibnamefont {Bostwick}}, \bibinfo {author}
  {\bibfnamefont {Roland~J.}\ \bibnamefont {Koch}}, \bibinfo {author}
  {\bibfnamefont {S\o{}ren}\ \bibnamefont {Ulstrup}}, \ and\ \bibinfo {author}
  {\bibfnamefont {Jyoti}\ \bibnamefont {Katoch}},\ }\bibfield  {title}
  {\enquote {\bibinfo {title} {Momentum-resolved view of highly tunable
  many-body effects in a graphene/h{BN} field-effect device},}\ }\href
  {\doibase 10.1103/PhysRevB.101.201409} {\bibfield  {journal} {\bibinfo
  {journal} {Phys. Rev. B}\ }\textbf {\bibinfo {volume} {101}},\ \bibinfo
  {pages} {201409} (\bibinfo {year} {2020})}\BibitemShut {NoStop}%
\bibitem [{\citenamefont {Jones}\ \emph {et~al.}(2020)\citenamefont {Jones},
  \citenamefont {Muzzio}, \citenamefont {Majchrzak}, \citenamefont {Pakdel},
  \citenamefont {Curcio}, \citenamefont {Volckaert}, \citenamefont {Biswas},
  \citenamefont {Gobbo}, \citenamefont {Singh}, \citenamefont {Robinson},
  \citenamefont {Watanabe}, \citenamefont {Taniguchi}, \citenamefont {Kim},
  \citenamefont {Cacho}, \citenamefont {Lanata}, \citenamefont {Miwa},
  \citenamefont {Hofmann}, \citenamefont {Katoch},\ and\ \citenamefont
  {Ulstrup}}]{Jones2020}%
  \BibitemOpen
  \bibfield  {author} {\bibinfo {author} {\bibfnamefont {Alfred J.~H.}\
  \bibnamefont {Jones}}, \bibinfo {author} {\bibfnamefont {Ryan}\ \bibnamefont
  {Muzzio}}, \bibinfo {author} {\bibfnamefont {Paulina}\ \bibnamefont
  {Majchrzak}}, \bibinfo {author} {\bibfnamefont {Sahar}\ \bibnamefont
  {Pakdel}}, \bibinfo {author} {\bibfnamefont {Davide}\ \bibnamefont {Curcio}},
  \bibinfo {author} {\bibfnamefont {Klara}\ \bibnamefont {Volckaert}}, \bibinfo
  {author} {\bibfnamefont {Deepnarayan}\ \bibnamefont {Biswas}}, \bibinfo
  {author} {\bibfnamefont {Jacob}\ \bibnamefont {Gobbo}}, \bibinfo {author}
  {\bibfnamefont {Simranjeet}\ \bibnamefont {Singh}}, \bibinfo {author}
  {\bibfnamefont {Jeremy~T.}\ \bibnamefont {Robinson}}, \bibinfo {author}
  {\bibfnamefont {Kenji}\ \bibnamefont {Watanabe}}, \bibinfo {author}
  {\bibfnamefont {Takashi}\ \bibnamefont {Taniguchi}}, \bibinfo {author}
  {\bibfnamefont {Timur~K.}\ \bibnamefont {Kim}}, \bibinfo {author}
  {\bibfnamefont {Cephise}\ \bibnamefont {Cacho}}, \bibinfo {author}
  {\bibfnamefont {Nicola}\ \bibnamefont {Lanata}}, \bibinfo {author}
  {\bibfnamefont {Jill~A.}\ \bibnamefont {Miwa}}, \bibinfo {author}
  {\bibfnamefont {Philip}\ \bibnamefont {Hofmann}}, \bibinfo {author}
  {\bibfnamefont {Jyoti}\ \bibnamefont {Katoch}}, \ and\ \bibinfo {author}
  {\bibfnamefont {Søren}\ \bibnamefont {Ulstrup}},\ }\bibfield  {title}
  {\enquote {\bibinfo {title} {Observation of electrically tunable van hove
  singularities in twisted bilayer graphene from nanoarpes},}\ }\href {\doibase
  https://doi.org/10.1002/adma.202001656} {\bibfield  {journal} {\bibinfo
  {journal} {Advanced Materials}\ }\textbf {\bibinfo {volume} {32}},\ \bibinfo
  {pages} {2001656} (\bibinfo {year} {2020})}\BibitemShut {NoStop}%
\bibitem [{\citenamefont {Hofmann}(2021)}]{Hofmann:2021}%
  \BibitemOpen
  \bibfield  {author} {\bibinfo {author} {\bibfnamefont {Philip}\ \bibnamefont
  {Hofmann}},\ }\bibfield  {title} {\enquote {\bibinfo {title} {{Accessing the
  spectral function of in operando devices by angle-resolved photoemission
  spectroscopy}},}\ }\href {\doibase 10.1116/5.0038637} {\bibfield  {journal}
  {\bibinfo  {journal} {AVS Quantum Science}\ }\textbf {\bibinfo {volume} {3}}
  (\bibinfo {year} {2021}),\ 10.1116/5.0038637}\BibitemShut {NoStop}%
\bibitem [{\citenamefont {Dale}\ \emph {et~al.}(2022)\citenamefont {Dale},
  \citenamefont {Mori}, \citenamefont {Utama}, \citenamefont {Denlinger},
  \citenamefont {Stansbury}, \citenamefont {Fatuzzo}, \citenamefont {Zhao},
  \citenamefont {Lee}, \citenamefont {Taniguchi}, \citenamefont {Watanabe},
  \citenamefont {Jozwiak}, \citenamefont {Bostwick}, \citenamefont {Rotenberg},
  \citenamefont {Koch}, \citenamefont {Wang},\ and\ \citenamefont
  {Lanzara}}]{Dale:2022}%
  \BibitemOpen
  \bibfield  {author} {\bibinfo {author} {\bibfnamefont {Nicholas}\
  \bibnamefont {Dale}}, \bibinfo {author} {\bibfnamefont {Ryo}\ \bibnamefont
  {Mori}}, \bibinfo {author} {\bibfnamefont {M.~Iqbal~Bakti}\ \bibnamefont
  {Utama}}, \bibinfo {author} {\bibfnamefont {Jonathan~D.}\ \bibnamefont
  {Denlinger}}, \bibinfo {author} {\bibfnamefont {Conrad}\ \bibnamefont
  {Stansbury}}, \bibinfo {author} {\bibfnamefont {Claudia~G.}\ \bibnamefont
  {Fatuzzo}}, \bibinfo {author} {\bibfnamefont {Sihan}\ \bibnamefont {Zhao}},
  \bibinfo {author} {\bibfnamefont {Kyunghoon}\ \bibnamefont {Lee}}, \bibinfo
  {author} {\bibfnamefont {Takashi}\ \bibnamefont {Taniguchi}}, \bibinfo
  {author} {\bibfnamefont {Kenji}\ \bibnamefont {Watanabe}}, \bibinfo {author}
  {\bibfnamefont {Chris}\ \bibnamefont {Jozwiak}}, \bibinfo {author}
  {\bibfnamefont {Aaron}\ \bibnamefont {Bostwick}}, \bibinfo {author}
  {\bibfnamefont {Eli}\ \bibnamefont {Rotenberg}}, \bibinfo {author}
  {\bibfnamefont {Roland~J.}\ \bibnamefont {Koch}}, \bibinfo {author}
  {\bibfnamefont {Feng}\ \bibnamefont {Wang}}, \ and\ \bibinfo {author}
  {\bibfnamefont {Alessandra}\ \bibnamefont {Lanzara}},\ }\bibfield  {title}
  {\enquote {\bibinfo {title} {Correlation-driven electron-hole asymmetry in
  graphene field effect devices},}\ }\href {\doibase
  10.1038/s41535-021-00404-8} {\bibfield  {journal} {\bibinfo  {journal} {npj
  Quantum Materials}\ }\textbf {\bibinfo {volume} {7}},\ \bibinfo {pages} {9}
  (\bibinfo {year} {2022})}\BibitemShut {NoStop}%
\bibitem [{\citenamefont {Jones}\ \emph {et~al.}(2022)\citenamefont {Jones},
  \citenamefont {Gammelgaard}, \citenamefont {Sauer}, \citenamefont {Biswas},
  \citenamefont {Koch}, \citenamefont {Jozwiak}, \citenamefont {Rotenberg},
  \citenamefont {Bostwick}, \citenamefont {Watanabe}, \citenamefont
  {Taniguchi}, \citenamefont {Dean}, \citenamefont {Jauho}, \citenamefont
  {B{\o}ggild}, \citenamefont {Pedersen}, \citenamefont {Jessen},\ and\
  \citenamefont {Ulstrup}}]{Jones:2022}%
  \BibitemOpen
  \bibfield  {author} {\bibinfo {author} {\bibfnamefont {Alfred J.~H.}\
  \bibnamefont {Jones}}, \bibinfo {author} {\bibfnamefont {Lene}\ \bibnamefont
  {Gammelgaard}}, \bibinfo {author} {\bibfnamefont {Mikkel~O.}\ \bibnamefont
  {Sauer}}, \bibinfo {author} {\bibfnamefont {Deepnarayan}\ \bibnamefont
  {Biswas}}, \bibinfo {author} {\bibfnamefont {Roland~J.}\ \bibnamefont
  {Koch}}, \bibinfo {author} {\bibfnamefont {Chris}\ \bibnamefont {Jozwiak}},
  \bibinfo {author} {\bibfnamefont {Eli}\ \bibnamefont {Rotenberg}}, \bibinfo
  {author} {\bibfnamefont {Aaron}\ \bibnamefont {Bostwick}}, \bibinfo {author}
  {\bibfnamefont {Kenji}\ \bibnamefont {Watanabe}}, \bibinfo {author}
  {\bibfnamefont {Takashi}\ \bibnamefont {Taniguchi}}, \bibinfo {author}
  {\bibfnamefont {Cory~R.}\ \bibnamefont {Dean}}, \bibinfo {author}
  {\bibfnamefont {Antti-Pekka}\ \bibnamefont {Jauho}}, \bibinfo {author}
  {\bibfnamefont {Peter}\ \bibnamefont {B{\o}ggild}}, \bibinfo {author}
  {\bibfnamefont {Thomas~G.}\ \bibnamefont {Pedersen}}, \bibinfo {author}
  {\bibfnamefont {Bjarke~S.}\ \bibnamefont {Jessen}}, \ and\ \bibinfo {author}
  {\bibfnamefont {S{\o}ren}\ \bibnamefont {Ulstrup}},\ }\bibfield  {title}
  {\enquote {\bibinfo {title} {Nanoscale view of engineered massive dirac
  quasiparticles in lithographic superstructures},}\ }\href {\doibase
  10.1021/acsnano.2c08929} {\bibfield  {journal} {\bibinfo  {journal} {ACS
  Nano}\ }\textbf {\bibinfo {volume} {16}},\ \bibinfo {pages} {19354--19362}
  (\bibinfo {year} {2022})}\BibitemShut {NoStop}%
\bibitem [{\citenamefont {Nunn}\ \emph {et~al.}(2023)\citenamefont {Nunn},
  \citenamefont {McEllistrim}, \citenamefont {Weston}, \citenamefont
  {Garcia-Ruiz}, \citenamefont {Watson}, \citenamefont {Mucha-Kruczynski},
  \citenamefont {Cacho}, \citenamefont {Gorbachev}, \citenamefont {Fal'ko},\
  and\ \citenamefont {Wilson}}]{Nunn:2023}%
  \BibitemOpen
  \bibfield  {author} {\bibinfo {author} {\bibfnamefont {James~E.}\
  \bibnamefont {Nunn}}, \bibinfo {author} {\bibfnamefont {Andrew}\ \bibnamefont
  {McEllistrim}}, \bibinfo {author} {\bibfnamefont {Astrid}\ \bibnamefont
  {Weston}}, \bibinfo {author} {\bibfnamefont {Aitor}\ \bibnamefont
  {Garcia-Ruiz}}, \bibinfo {author} {\bibfnamefont {Matthew~D.}\ \bibnamefont
  {Watson}}, \bibinfo {author} {\bibfnamefont {Marcin}\ \bibnamefont
  {Mucha-Kruczynski}}, \bibinfo {author} {\bibfnamefont {Cephise}\ \bibnamefont
  {Cacho}}, \bibinfo {author} {\bibfnamefont {Roman~V.}\ \bibnamefont
  {Gorbachev}}, \bibinfo {author} {\bibfnamefont {Vladimir~I.}\ \bibnamefont
  {Fal'ko}}, \ and\ \bibinfo {author} {\bibfnamefont {Neil~R.}\ \bibnamefont
  {Wilson}},\ }\bibfield  {title} {\enquote {\bibinfo {title} {Arpes signatures
  of few-layer twistronic graphenes},}\ }\href {\doibase
  10.1021/acs.nanolett.3c01173} {\bibfield  {journal} {\bibinfo  {journal}
  {Nano Letters}\ }\textbf {\bibinfo {volume} {23}},\ \bibinfo {pages}
  {5201--5208} (\bibinfo {year} {2023})}\BibitemShut {NoStop}%
\bibitem [{\citenamefont {Dale}\ \emph {et~al.}(2023)\citenamefont {Dale},
  \citenamefont {Utama}, \citenamefont {Lee}, \citenamefont {Leconte},
  \citenamefont {Zhao}, \citenamefont {Lee}, \citenamefont {Taniguchi},
  \citenamefont {Watanabe}, \citenamefont {Jozwiak}, \citenamefont {Bostwick},
  \citenamefont {Rotenberg}, \citenamefont {Koch}, \citenamefont {Jung},
  \citenamefont {Wang},\ and\ \citenamefont {Lanzara}}]{nDale:2023}%
  \BibitemOpen
  \bibfield  {author} {\bibinfo {author} {\bibfnamefont {Nicholas}\
  \bibnamefont {Dale}}, \bibinfo {author} {\bibfnamefont {M.~Iqbal~Bakti}\
  \bibnamefont {Utama}}, \bibinfo {author} {\bibfnamefont {Dongkyu}\
  \bibnamefont {Lee}}, \bibinfo {author} {\bibfnamefont {Nicolas}\ \bibnamefont
  {Leconte}}, \bibinfo {author} {\bibfnamefont {Sihan}\ \bibnamefont {Zhao}},
  \bibinfo {author} {\bibfnamefont {Kyunghoon}\ \bibnamefont {Lee}}, \bibinfo
  {author} {\bibfnamefont {Takashi}\ \bibnamefont {Taniguchi}}, \bibinfo
  {author} {\bibfnamefont {Kenji}\ \bibnamefont {Watanabe}}, \bibinfo {author}
  {\bibfnamefont {Chris}\ \bibnamefont {Jozwiak}}, \bibinfo {author}
  {\bibfnamefont {Aaron}\ \bibnamefont {Bostwick}}, \bibinfo {author}
  {\bibfnamefont {Eli}\ \bibnamefont {Rotenberg}}, \bibinfo {author}
  {\bibfnamefont {Roland~J.}\ \bibnamefont {Koch}}, \bibinfo {author}
  {\bibfnamefont {Jeil}\ \bibnamefont {Jung}}, \bibinfo {author} {\bibfnamefont
  {Feng}\ \bibnamefont {Wang}}, \ and\ \bibinfo {author} {\bibfnamefont
  {Alessandra}\ \bibnamefont {Lanzara}},\ }\bibfield  {title} {\enquote
  {\bibinfo {title} {Layer-dependent interaction effects in the electronic
  structure of twisted bilayer graphene devices},}\ }\bibfield  {booktitle}
  {\emph {\bibinfo {booktitle} {Nano Letters}},\ }\href {\doibase
  10.1021/acs.nanolett.3c00253} {\bibfield  {journal} {\bibinfo  {journal}
  {Nano Letters}\ }\textbf {\bibinfo {volume} {23}},\ \bibinfo {pages}
  {6799--6806} (\bibinfo {year} {2023})}\BibitemShut {NoStop}%
\bibitem [{\citenamefont {Ju}\ \emph {et~al.}(2014)\citenamefont {Ju},
  \citenamefont {Velasco}, \citenamefont {Huang}, \citenamefont {Kahn},
  \citenamefont {Nosiglia}, \citenamefont {Tsai}, \citenamefont {Yang},
  \citenamefont {Taniguchi}, \citenamefont {Watanabe}, \citenamefont {Zhang},
  \citenamefont {Zhang}, \citenamefont {Crommie}, \citenamefont {Zettl},\ and\
  \citenamefont {Wang}}]{Velasco:2014}%
  \BibitemOpen
  \bibfield  {author} {\bibinfo {author} {\bibfnamefont {L.}~\bibnamefont
  {Ju}}, \bibinfo {author} {\bibfnamefont {J.}~\bibnamefont {Velasco}},
  \bibinfo {author} {\bibfnamefont {E.}~\bibnamefont {Huang}}, \bibinfo
  {author} {\bibfnamefont {S.}~\bibnamefont {Kahn}}, \bibinfo {author}
  {\bibfnamefont {C.}~\bibnamefont {Nosiglia}}, \bibinfo {author}
  {\bibfnamefont {Hsin-Zon}\ \bibnamefont {Tsai}}, \bibinfo {author}
  {\bibfnamefont {W.}~\bibnamefont {Yang}}, \bibinfo {author} {\bibfnamefont
  {T.}~\bibnamefont {Taniguchi}}, \bibinfo {author} {\bibfnamefont
  {K.}~\bibnamefont {Watanabe}}, \bibinfo {author} {\bibfnamefont
  {Y.}~\bibnamefont {Zhang}}, \bibinfo {author} {\bibfnamefont
  {G.}~\bibnamefont {Zhang}}, \bibinfo {author} {\bibfnamefont
  {M.}~\bibnamefont {Crommie}}, \bibinfo {author} {\bibfnamefont
  {A.}~\bibnamefont {Zettl}}, \ and\ \bibinfo {author} {\bibfnamefont
  {F.}~\bibnamefont {Wang}},\ }\bibfield  {title} {\enquote {\bibinfo {title}
  {Photoinduced doping in heterostructures of graphene and boron nitride},}\
  }\href {\doibase 10.1038/nnano.2014.60} {\bibfield  {journal} {\bibinfo
  {journal} {Nature Nanotechnology}\ }\textbf {\bibinfo {volume} {9}},\
  \bibinfo {pages} {348--352} (\bibinfo {year} {2014})}\BibitemShut {NoStop}%
\bibitem [{\citenamefont {Bianchi}\ \emph {et~al.}(2023)\citenamefont
  {Bianchi}, \citenamefont {Hofmann}, \citenamefont {Hoffmann}, \citenamefont
  {Jones}, \citenamefont {Li}, \citenamefont {Miwa}, \citenamefont {M{\o}ller},
  \citenamefont {Nielsen}, \citenamefont {Thomsen}, \citenamefont {Ulstrup},\
  and\ \citenamefont {Worm}}]{Bianchi:2023}%
  \BibitemOpen
  \bibfield  {author} {\bibinfo {author} {\bibfnamefont {Marco}\ \bibnamefont
  {Bianchi}}, \bibinfo {author} {\bibfnamefont {Philip}\ \bibnamefont
  {Hofmann}}, \bibinfo {author} {\bibfnamefont {S{\o}ren~V.}\ \bibnamefont
  {Hoffmann}}, \bibinfo {author} {\bibfnamefont {Nykola~C.}\ \bibnamefont
  {Jones}}, \bibinfo {author} {\bibfnamefont {Zheshen}\ \bibnamefont {Li}},
  \bibinfo {author} {\bibfnamefont {Jill~A.}\ \bibnamefont {Miwa}}, \bibinfo
  {author} {\bibfnamefont {S{\o}ren~P.}\ \bibnamefont {M{\o}ller}}, \bibinfo
  {author} {\bibfnamefont {J{\o}rgen~S.}\ \bibnamefont {Nielsen}}, \bibinfo
  {author} {\bibfnamefont {Heine~D.}\ \bibnamefont {Thomsen}}, \bibinfo
  {author} {\bibfnamefont {S{\o}ren}\ \bibnamefont {Ulstrup}}, \ and\ \bibinfo
  {author} {\bibfnamefont {Torben}\ \bibnamefont {Worm}},\ }\bibfield  {title}
  {\enquote {\bibinfo {title} {Status and strategy at isa, centre for storage
  ring facilities, aarhus university, denmark},}\ }\href {\doibase
  10.1140/epjp/s13360-023-03748-1} {\bibfield  {journal} {\bibinfo  {journal}
  {The European Physical Journal Plus}\ }\textbf {\bibinfo {volume} {138}},\
  \bibinfo {pages} {132} (\bibinfo {year} {2023})}\BibitemShut {NoStop}%
\bibitem [{\citenamefont {Volckaert}\ \emph {et~al.}()\citenamefont
  {Volckaert}, \citenamefont {Majchrzak}, \citenamefont {Biswas}, \citenamefont
  {Jones}, \citenamefont {Bianchi}, \citenamefont {Jiang}, \citenamefont
  {Dubourg}, \citenamefont {Stenshøj}, \citenamefont {Jensen}, \citenamefont
  {Jones}, \citenamefont {Hoffmann}, \citenamefont {Mi}, \citenamefont
  {Bremholm}, \citenamefont {Pan}, \citenamefont {Chen}, \citenamefont
  {Hofmann}, \citenamefont {Miwa},\ and\ \citenamefont
  {Ulstrup}}]{Volckaert:2023}%
  \BibitemOpen
  \bibfield  {author} {\bibinfo {author} {\bibfnamefont {Klara}\ \bibnamefont
  {Volckaert}}, \bibinfo {author} {\bibfnamefont {Paulina}\ \bibnamefont
  {Majchrzak}}, \bibinfo {author} {\bibfnamefont {Deepnarayan}\ \bibnamefont
  {Biswas}}, \bibinfo {author} {\bibfnamefont {Alfred J.~H.}\ \bibnamefont
  {Jones}}, \bibinfo {author} {\bibfnamefont {Marco}\ \bibnamefont {Bianchi}},
  \bibinfo {author} {\bibfnamefont {Zhihao}\ \bibnamefont {Jiang}}, \bibinfo
  {author} {\bibfnamefont {Raphaël}\ \bibnamefont {Dubourg}}, \bibinfo
  {author} {\bibfnamefont {Rasmus~Ørnekoll}\ \bibnamefont {Stenshøj}},
  \bibinfo {author} {\bibfnamefont {Mads~Lykke}\ \bibnamefont {Jensen}},
  \bibinfo {author} {\bibfnamefont {Nykola~C.}\ \bibnamefont {Jones}}, \bibinfo
  {author} {\bibfnamefont {Søren~V.}\ \bibnamefont {Hoffmann}}, \bibinfo
  {author} {\bibfnamefont {Jian-Li}\ \bibnamefont {Mi}}, \bibinfo {author}
  {\bibfnamefont {Martin}\ \bibnamefont {Bremholm}}, \bibinfo {author}
  {\bibfnamefont {Xing-Chen}\ \bibnamefont {Pan}}, \bibinfo {author}
  {\bibfnamefont {Yong~P.}\ \bibnamefont {Chen}}, \bibinfo {author}
  {\bibfnamefont {Philip}\ \bibnamefont {Hofmann}}, \bibinfo {author}
  {\bibfnamefont {Jill~A.}\ \bibnamefont {Miwa}}, \ and\ \bibinfo {author}
  {\bibfnamefont {Søren}\ \bibnamefont {Ulstrup}},\ }\bibfield  {title}
  {\enquote {\bibinfo {title} {Surface electronic structure engineering of
  manganese bismuth tellurides guided by micro-focused angle-resolved
  photoemission},}\ }\href {\doibase https://doi.org/10.1002/adma.202301907}
  {\bibfield  {journal} {\bibinfo  {journal} {Advanced Materials}\ }\textbf
  {\bibinfo {volume} {n/a}},\ \bibinfo {pages} {2301907}},\ \Eprint
  {http://arxiv.org/abs/https://onlinelibrary.wiley.com/doi/pdf/10.1002/adma.202301907}
  {https://onlinelibrary.wiley.com/doi/pdf/10.1002/adma.202301907} \BibitemShut
  {NoStop}%
\bibitem [{\citenamefont {Koch}\ \emph
  {et~al.}(2018{\natexlab{a}})\citenamefont {Koch}, \citenamefont {Katoch},
  \citenamefont {Moser}, \citenamefont {Schwarz}, \citenamefont {Kawakami},
  \citenamefont {Bostwick}, \citenamefont {Rotenberg}, \citenamefont
  {Jozwiak},\ and\ \citenamefont {Ulstrup}}]{Koch:2018ab}%
  \BibitemOpen
  \bibfield  {author} {\bibinfo {author} {\bibfnamefont {Roland~J.}\
  \bibnamefont {Koch}}, \bibinfo {author} {\bibfnamefont {Jyoti}\ \bibnamefont
  {Katoch}}, \bibinfo {author} {\bibfnamefont {Simon}\ \bibnamefont {Moser}},
  \bibinfo {author} {\bibfnamefont {Daniel}\ \bibnamefont {Schwarz}}, \bibinfo
  {author} {\bibfnamefont {Roland~K.}\ \bibnamefont {Kawakami}}, \bibinfo
  {author} {\bibfnamefont {Aaron}\ \bibnamefont {Bostwick}}, \bibinfo {author}
  {\bibfnamefont {Eli}\ \bibnamefont {Rotenberg}}, \bibinfo {author}
  {\bibfnamefont {Chris}\ \bibnamefont {Jozwiak}}, \ and\ \bibinfo {author}
  {\bibfnamefont {S\o{}ren}\ \bibnamefont {Ulstrup}},\ }\bibfield  {title}
  {\enquote {\bibinfo {title} {Electronic structure of exfoliated and epitaxial
  hexagonal boron nitride},}\ }\href {\doibase
  10.1103/PhysRevMaterials.2.074006} {\bibfield  {journal} {\bibinfo  {journal}
  {Phys. Rev. Materials}\ }\textbf {\bibinfo {volume} {2}},\ \bibinfo {pages}
  {074006} (\bibinfo {year} {2018}{\natexlab{a}})}\BibitemShut {NoStop}%
\bibitem [{\citenamefont {Koch}\ \emph
  {et~al.}(2018{\natexlab{b}})\citenamefont {Koch}, \citenamefont {Jozwiak},
  \citenamefont {Bostwick}, \citenamefont {Stripe}, \citenamefont {Cordier},
  \citenamefont {Hussain}, \citenamefont {Yun},\ and\ \citenamefont
  {Rotenberg}}]{Koch2018}%
  \BibitemOpen
  \bibfield  {author} {\bibinfo {author} {\bibfnamefont {Roland~J.}\
  \bibnamefont {Koch}}, \bibinfo {author} {\bibfnamefont {Chris}\ \bibnamefont
  {Jozwiak}}, \bibinfo {author} {\bibfnamefont {Aaron}\ \bibnamefont
  {Bostwick}}, \bibinfo {author} {\bibfnamefont {Benjamin}\ \bibnamefont
  {Stripe}}, \bibinfo {author} {\bibfnamefont {Mark}\ \bibnamefont {Cordier}},
  \bibinfo {author} {\bibfnamefont {Zahid}\ \bibnamefont {Hussain}}, \bibinfo
  {author} {\bibfnamefont {Wenbing}\ \bibnamefont {Yun}}, \ and\ \bibinfo
  {author} {\bibfnamefont {Eli}\ \bibnamefont {Rotenberg}},\ }\bibfield
  {title} {\enquote {\bibinfo {title} {Nano focusing of soft x-rays by a new
  capillary mirror optic},}\ }\href {\doibase 10.1080/08940886.2018.1483660}
  {\bibfield  {journal} {\bibinfo  {journal} {Synchrotron Radiation News}\
  }\textbf {\bibinfo {volume} {31}},\ \bibinfo {pages} {50--52} (\bibinfo
  {year} {2018}{\natexlab{b}})}\BibitemShut {NoStop}%
\bibitem [{\citenamefont {Ulstrup}\ \emph {et~al.}(2020)\citenamefont
  {Ulstrup}, \citenamefont {Koch}, \citenamefont {Singh}, \citenamefont
  {McCreary}, \citenamefont {Jonker}, \citenamefont {Robinson}, \citenamefont
  {Jozwiak}, \citenamefont {Rotenberg}, \citenamefont {Bostwick}, \citenamefont
  {Katoch},\ and\ \citenamefont {Miwa}}]{Ulstrup2020}%
  \BibitemOpen
  \bibfield  {author} {\bibinfo {author} {\bibfnamefont {S{\o}ren}\
  \bibnamefont {Ulstrup}}, \bibinfo {author} {\bibfnamefont {Roland~J.}\
  \bibnamefont {Koch}}, \bibinfo {author} {\bibfnamefont {Simranjeet}\
  \bibnamefont {Singh}}, \bibinfo {author} {\bibfnamefont {Kathleen~M.}\
  \bibnamefont {McCreary}}, \bibinfo {author} {\bibfnamefont {Berend~T.}\
  \bibnamefont {Jonker}}, \bibinfo {author} {\bibfnamefont {Jeremy~T.}\
  \bibnamefont {Robinson}}, \bibinfo {author} {\bibfnamefont {Chris}\
  \bibnamefont {Jozwiak}}, \bibinfo {author} {\bibfnamefont {Eli}\ \bibnamefont
  {Rotenberg}}, \bibinfo {author} {\bibfnamefont {Aaron}\ \bibnamefont
  {Bostwick}}, \bibinfo {author} {\bibfnamefont {Jyoti}\ \bibnamefont
  {Katoch}}, \ and\ \bibinfo {author} {\bibfnamefont {Jill~A.}\ \bibnamefont
  {Miwa}},\ }\bibfield  {title} {\enquote {\bibinfo {title} {Direct observation
  of minibands in a twisted graphene/ws2 bilayer},}\ }\href {\doibase
  10.1126/sciadv.aay6104} {\bibfield  {journal} {\bibinfo  {journal} {Science
  Advances}\ }\textbf {\bibinfo {volume} {6}},\ \bibinfo {pages} {eaay6104}
  (\bibinfo {year} {2020})}\BibitemShut {NoStop}%
\bibitem [{\citenamefont {Morozov}\ \emph {et~al.}(2008)\citenamefont
  {Morozov}, \citenamefont {Novoselov}, \citenamefont {Katsnelson},
  \citenamefont {Schedin}, \citenamefont {Elias}, \citenamefont {Jaszczak},\
  and\ \citenamefont {Geim}}]{Morozov:2008}%
  \BibitemOpen
  \bibfield  {author} {\bibinfo {author} {\bibfnamefont {S.~V.}\ \bibnamefont
  {Morozov}}, \bibinfo {author} {\bibfnamefont {K.~S.}\ \bibnamefont
  {Novoselov}}, \bibinfo {author} {\bibfnamefont {M.~I.}\ \bibnamefont
  {Katsnelson}}, \bibinfo {author} {\bibfnamefont {F.}~\bibnamefont {Schedin}},
  \bibinfo {author} {\bibfnamefont {D.~C.}\ \bibnamefont {Elias}}, \bibinfo
  {author} {\bibfnamefont {J.~A.}\ \bibnamefont {Jaszczak}}, \ and\ \bibinfo
  {author} {\bibfnamefont {A.~K.}\ \bibnamefont {Geim}},\ }\bibfield  {title}
  {\enquote {\bibinfo {title} {Giant intrinsic carrier mobilities in graphene
  and its bilayer},}\ }\href {\doibase 10.1103/PhysRevLett.100.016602}
  {\bibfield  {journal} {\bibinfo  {journal} {Phys. Rev. Lett.}\ }\textbf
  {\bibinfo {volume} {100}},\ \bibinfo {pages} {016602} (\bibinfo {year}
  {2008})}\BibitemShut {NoStop}%
\bibitem [{\citenamefont {Yang}\ \emph {et~al.}(2011)\citenamefont {Yang},
  \citenamefont {Balakrishnan}, \citenamefont {Volmer}, \citenamefont {Avsar},
  \citenamefont {Jaiswal}, \citenamefont {Samm}, \citenamefont {Ali},
  \citenamefont {Pachoud}, \citenamefont {Zeng}, \citenamefont {Popinciuc},
  \citenamefont {G\"untherodt}, \citenamefont {Beschoten},\ and\ \citenamefont
  {\"Ozyilmaz}}]{Yang:2011}%
  \BibitemOpen
  \bibfield  {author} {\bibinfo {author} {\bibfnamefont {T.-Y.}\ \bibnamefont
  {Yang}}, \bibinfo {author} {\bibfnamefont {J.}~\bibnamefont {Balakrishnan}},
  \bibinfo {author} {\bibfnamefont {F.}~\bibnamefont {Volmer}}, \bibinfo
  {author} {\bibfnamefont {A.}~\bibnamefont {Avsar}}, \bibinfo {author}
  {\bibfnamefont {M.}~\bibnamefont {Jaiswal}}, \bibinfo {author} {\bibfnamefont
  {J.}~\bibnamefont {Samm}}, \bibinfo {author} {\bibfnamefont {S.~R.}\
  \bibnamefont {Ali}}, \bibinfo {author} {\bibfnamefont {A.}~\bibnamefont
  {Pachoud}}, \bibinfo {author} {\bibfnamefont {M.}~\bibnamefont {Zeng}},
  \bibinfo {author} {\bibfnamefont {M.}~\bibnamefont {Popinciuc}}, \bibinfo
  {author} {\bibfnamefont {G.}~\bibnamefont {G\"untherodt}}, \bibinfo {author}
  {\bibfnamefont {B.}~\bibnamefont {Beschoten}}, \ and\ \bibinfo {author}
  {\bibfnamefont {B.}~\bibnamefont {\"Ozyilmaz}},\ }\bibfield  {title}
  {\enquote {\bibinfo {title} {Observation of long spin-relaxation times in
  bilayer graphene at room temperature},}\ }\href {\doibase
  10.1103/PhysRevLett.107.047206} {\bibfield  {journal} {\bibinfo  {journal}
  {Phys. Rev. Lett.}\ }\textbf {\bibinfo {volume} {107}},\ \bibinfo {pages}
  {047206} (\bibinfo {year} {2011})}\BibitemShut {NoStop}%
\bibitem [{\citenamefont {Maassen}\ \emph {et~al.}(2011)\citenamefont
  {Maassen}, \citenamefont {Dejene}, \citenamefont {Guimar\~aes}, \citenamefont
  {J\'ozsa},\ and\ \citenamefont {van Wees}}]{Maassen:2011}%
  \BibitemOpen
  \bibfield  {author} {\bibinfo {author} {\bibfnamefont {T.}~\bibnamefont
  {Maassen}}, \bibinfo {author} {\bibfnamefont {F.~K.}\ \bibnamefont {Dejene}},
  \bibinfo {author} {\bibfnamefont {M.~H.~D.}\ \bibnamefont {Guimar\~aes}},
  \bibinfo {author} {\bibfnamefont {C.}~\bibnamefont {J\'ozsa}}, \ and\
  \bibinfo {author} {\bibfnamefont {B.~J.}\ \bibnamefont {van Wees}},\
  }\bibfield  {title} {\enquote {\bibinfo {title} {Comparison between charge
  and spin transport in few-layer graphene},}\ }\href {\doibase
  10.1103/PhysRevB.83.115410} {\bibfield  {journal} {\bibinfo  {journal} {Phys.
  Rev. B}\ }\textbf {\bibinfo {volume} {83}},\ \bibinfo {pages} {115410}
  (\bibinfo {year} {2011})}\BibitemShut {NoStop}%
\bibitem [{\citenamefont {Gierz}\ \emph {et~al.}(2011)\citenamefont {Gierz},
  \citenamefont {Henk}, \citenamefont {H\"ochst}, \citenamefont {Ast},\ and\
  \citenamefont {Kern}}]{Gierz2011.PhysRevB.83.121408}%
  \BibitemOpen
  \bibfield  {author} {\bibinfo {author} {\bibfnamefont {Isabella}\
  \bibnamefont {Gierz}}, \bibinfo {author} {\bibfnamefont {J\"urgen}\
  \bibnamefont {Henk}}, \bibinfo {author} {\bibfnamefont {Hartmut}\
  \bibnamefont {H\"ochst}}, \bibinfo {author} {\bibfnamefont {Christian~R.}\
  \bibnamefont {Ast}}, \ and\ \bibinfo {author} {\bibfnamefont {Klaus}\
  \bibnamefont {Kern}},\ }\bibfield  {title} {\enquote {\bibinfo {title}
  {Illuminating the dark corridor in graphene: Polarization dependence of
  angle-resolved photoemission spectroscopy on graphene},}\ }\href {\doibase
  10.1103/PhysRevB.83.121408} {\bibfield  {journal} {\bibinfo  {journal} {Phys.
  Rev. B}\ }\textbf {\bibinfo {volume} {83}},\ \bibinfo {pages} {121408}
  (\bibinfo {year} {2011})}\BibitemShut {NoStop}%
\bibitem [{\citenamefont {Zhang}\ \emph {et~al.}(2011)\citenamefont {Zhang},
  \citenamefont {Richard}, \citenamefont {Qian}, \citenamefont {Xu},
  \citenamefont {Dai},\ and\ \citenamefont {Ding}}]{CurvM:2011}%
  \BibitemOpen
  \bibfield  {author} {\bibinfo {author} {\bibfnamefont {P.}~\bibnamefont
  {Zhang}}, \bibinfo {author} {\bibfnamefont {P.}~\bibnamefont {Richard}},
  \bibinfo {author} {\bibfnamefont {T.}~\bibnamefont {Qian}}, \bibinfo {author}
  {\bibfnamefont {Y.-M.}\ \bibnamefont {Xu}}, \bibinfo {author} {\bibfnamefont
  {X.}~\bibnamefont {Dai}}, \ and\ \bibinfo {author} {\bibfnamefont
  {H.}~\bibnamefont {Ding}},\ }\bibfield  {title} {\enquote {\bibinfo {title}
  {A precise method for visualizing dispersive features in image plots},}\
  }\href {\doibase 10.1063/1.3585113} {\bibfield  {journal} {\bibinfo
  {journal} {Review of Scientific Instruments}\ }\textbf {\bibinfo {volume}
  {82}},\ \bibinfo {pages} {043712} (\bibinfo {year} {2011})}\BibitemShut
  {NoStop}%
\bibitem [{\citenamefont {Ohta}\ \emph {et~al.}(2006)\citenamefont {Ohta},
  \citenamefont {Bostwick}, \citenamefont {Seyller}, \citenamefont {Horn},\
  and\ \citenamefont {Rotenberg}}]{Ohta:2006}%
  \BibitemOpen
  \bibfield  {author} {\bibinfo {author} {\bibfnamefont {Taisuke}\ \bibnamefont
  {Ohta}}, \bibinfo {author} {\bibfnamefont {Aaron}\ \bibnamefont {Bostwick}},
  \bibinfo {author} {\bibfnamefont {Thomas}\ \bibnamefont {Seyller}}, \bibinfo
  {author} {\bibfnamefont {Karsten}\ \bibnamefont {Horn}}, \ and\ \bibinfo
  {author} {\bibfnamefont {Eli}\ \bibnamefont {Rotenberg}},\ }\bibfield
  {title} {\enquote {\bibinfo {title} {Controlling the electronic structure of
  bilayer graphene},}\ }\href {\doibase 10.1126/science.1130681} {\bibfield
  {journal} {\bibinfo  {journal} {Science}\ }\textbf {\bibinfo {volume}
  {313}},\ \bibinfo {pages} {951--954} (\bibinfo {year} {2006})}\BibitemShut
  {NoStop}%
\end{thebibliography}
\end{document}